\documentclass[english]{article}
\usepackage[T1]{fontenc}
\usepackage[latin9]{inputenc}
\usepackage{geometry}
\usepackage{color}
\usepackage{graphicx}
\usepackage{subcaption}
\usepackage{amssymb}
\usepackage{epstopdf}
\usepackage{amsmath}
\usepackage{caption}
\newcommand{\comments}[1]{}   

\makeatletter

\usepackage{indentfirst}
\usepackage{enumerate}

\makeatother

\comments{
\documentclass[11pt,a4paper,final]{iopart}
\pdfoutput=1
\usepackage{graphicx}
\usepackage{cite}
\usepackage{bigfoot}
\usepackage{color}
\usepackage[dvipsnames]{xcolor}
\usepackage[breaklinks=true,colorlinks=true,linkcolor=blue,urlcolor=blue,citecolor=blue,pdfencoding=auto, psdextra]{hyperref}
\usepackage{amsfonts, amsmath, dsfont}
\usepackage{empheq}
\usepackage{subcaption}
\usepackage{amssymb}
\usepackage{epstopdf}
\usepackage{amsmath}
\usepackage{caption}
\newcommand{\comments}[1]{}   

\usepackage{indentfirst}
\usepackage{enumerate}

\makeatother
}

\begin{document}

\title{Non equilibrium dynamics of isolated disordered systems: \\
the classical Hamiltonian p-spin model}

\author{Leticia F. Cugliandolo$^1$, Gustavo S. Lozano$^2$ and Nicol\'as Nessi$^2$\\
$^1$  Sorbonne  Universit\'es,  Universit\'e  Pierre  et  Marie  Curie --- Paris  6, \\
Laboratoire  de  Physique  Th\'eorique  et  Hautes  Energies, \\
4,  Place Jussieu, 75252 Paris Cedex 05, France\\
$^2$ Departamento de F\'{\i}sica, \\
FCEYN Universidad de Buenos Aires
\& IFIBA CONICET,\\
Pabell\'on 1 Ciudad Universitaria, 1428 Buenos Aires, Argentina
}

\maketitle

\abstract{
We study the dynamics of a classical disordered macroscopic model completely isolated from the
environment reproducing, in a classical setting, the ``quantum quench'' protocol. We show that,
depending on the pre and post quench parameters the system approaches equilibrium, succeeding to
act as a bath on itself, or remains out of equilibrium, in two different ways.
In one of the latter, the system stays confined in a metastable state in which it
undergoes stationary dynamics characterised by a single temperature. In the other, the system
ages and its dynamics are characterised by two temperatures associated to observations made at short and long time differences
(high and low frequencies).
The parameter dependence of the asymptotic states is rationalised in terms of a
dynamic phase diagram with one equilibrium and two out of equilibrium phases.
Aspects of pre-thermalisation are observed and discussed. Similarities and differences with the dynamics
of the dissipative model are also explained.
}

\newpage

\tableofcontents

\newpage

\section{Introduction}
\label{sec:introduction}

The dynamics of isolated many-body
quantum systems, and especially the search of a statistical description
of their asymptotic evolution~\cite{Polkovnikov10}, are currently receiving much attention.
One of the motivations to study these issues theoretically is the
practical realisation of quenches of isolated  ultra-cold atoms trapped in optical lattices~\cite{Bloch08}.
Another reason for the interest in these questions, is the recently proposed many-body localisation phenomenon in
systems with quenched disorder~\cite{BaAlAl06}.

A {\it quantum} quench is the protocol whereby the ground state
(or, more generally, a mixed state) of a Hamiltonian $H_0$ is unitarily
evolved with a different Hamiltonian $H$. A similar procedure can also be realised in
a {\it classical}  setting. It amounts to evolve an isolated system initially prepared
in the equilibrium state (or, more generally, a metastable state) of a Hamiltonian $H_0$
with a Hamiltonian $H$ with the same form but different parameters. The latter problem has not received as much attention
as the former although we will show that it raises very similar questions as its quantum partner.

The most natural question to pose in the context of quenches of isolated classical or quantum
many-body physics is whether the system is able to provide a
bath for itself, allowing it to reach equilibration thanks  to the interactions.
In other words, the question is whether the system attains, in the long time limit, a stationary state of Gibbs-Boltzmann
thermal kind.

Two classes of quantum systems in which the answer to this question
is negative have already been identified.
These are {\it integrable systems}, with a macroscopic number of conserved quasi-local quantities~\cite{CaEsMu16},
and
{\it many-body localised quantum systems},
which
remain localised in states that are close to their initial conditions
forced by the frozen randomness~\cite{NaHu15,AlVo15}.
The non-ergodic behaviour of these systems
is expected to be  destabilised by the coupling to an external environment acting as a thermal bath~\cite{Huse}.

Our aim is to prove that there exists another class of isolated non-ergodic models that are not able to act as baths for themselves.
These are frustrated models with a complex free-energy landscapes that include, for wide ranges of variation of their parameters,
fully trapping regions. For the sake of clarity, and with the aim of distinguishing
effects that may be of unique quantum origin from features that are just due to the isolation of the model
and/or the peculiar character of the interactions, we start by treating a classical model.

In this paper we start a series of studies of the quench dynamics of isolated classical and quantum interacting disordered models
of mean-field kind, that is to say, models with $N$ fully-connected variables, endowed with a quenched random potential and kinetic energy.  Their choice
is motivated by the fact that their equilibrium, metastable and dissipative dynamics are very well understood
and realise the complex free-energy landscape able to render the dynamics non-ergodic. Concretely, they are
models with  random interactions between $p$ spins that have been extensively used as a mean-field description of the
glassy arrest and glassy phenomenology~\cite{KiTh87a,KiTh87b,KiWo87}.
In the infinite $p$ limit they become the random energy model~\cite{De80,GrMe84}, also used in the context of many-body
localisation studies~\cite{BaLaPaSc16a,BaLaPaSc16b}.

We show that, under different quenching conditions, the isolated dynamics of these {\it non-integrable} interacting systems
asymptotically approach:
\begin{enumerate}
\item[(i)]
A paramagnetic stationary state
described by a single temperature $T_f$, itself determined by
the final energy of the system $e_f$.
\item[(ii)]
A metastable state with stationary dynamics. Whether this steady state can or cannot be considered one of Gibbs-Boltzmann equilibrium, is  a subtle issue that we
will explain in the paper.
\item[(iii)]
A non-stationary ageing state described by two temperatures, $T_f$ and $T_{\rm eff}$, that are
also related to the final energy of the system $e_f$ and other non-trivial properties of it.
In this case, the system is clearly out of equilibrium.
\end{enumerate}

The impossibility to relax to thermal equilibrium is related to two prominent features of the potential energy landscape. In
case (ii),  the system reaches stationarity within one  (out of many) metastable state, which can be visualised as disconnected lakes in phase space.
Provided that the quench does not change the energy landscape too drastically, any trajectory starting from initial conditions inside a lake
will remain confined to it. The system will be unable to explore the whole phase space and, consequently,
it will not reach a state compatible with thermal equilibrium. The case (iii),
in which the system never reaches a stationary state, is related to the existence of the so-called threshold level, a region in phase space in which the potential energy is dominated by saddles points. Dynamics within the threshold level are characterised by slow relaxation and ageing.

We determine the dynamic phase diagram as a function of the pre and post quench parameters,
with dynamic transitions lines and phases that we characterise. The behaviour found is robust against the coupling to a bath.

The paper is organised as follows. In Sec.~\ref{sec:background} we introduce the model and we recall some of its main
properties. Section~\ref{sec:quenches} explains the quenches that we implement. In Sec.~\ref{sec:asymptotic} we present the
analytic developments and asymptotic results and in Sec.~\ref{sec:numerical} we show the
outcome of the numerical integration of the exact equations of motion. Section~\ref{sec:phase-diagram} presents the dynamic phase diagram.
Finally, in Sec.~\ref{sec:conclusions} we
present our conclusions and we briefly discuss our future projects.

\section{Background}
\label{sec:background}

\subsection{The Hamiltonian spherical $p$-spin model}

The $p$-spin model is a model with interactions between $p$ spins mediated by
quenched random couplings $J_{i_1 \dots i_p}$.
The potential energy is~\cite{KiTh87a,KiTh87b,KiWo87,De80,GrMe84}
\begin{eqnarray}
H_{\rm pot}[\{s_i\}] = - \frac{1}{p!} \sum^N_{i_1 \neq \dots \neq i_p} J_{i_1 \dots i_p} s_{i_1} \dots s_{i_p}
\; .
\label{eq:pspin-pot}
\end{eqnarray}
The coupling exchanges are independent identically distributed random variables taken from a Gaussian distribution with average and variance
\begin{equation}
[ J_{i_1 \dots i_p } ] =0
\; , \qquad\qquad
[ J^2_{i_1 \neq \dots \neq i_p } ] = \frac{J^2 p!}{2N^{p-1}}
\; .
\label{eq:disorder-statistics}
\end{equation}
The parameter $J$ characterises the width of the Gaussian.
In its standard spin-glass setting the spins are Ising variables. We will, instead, use continuous
variables, $-\sqrt{N} \leq s_i \leq \sqrt{N}$ with $i=1, \dots, N$, globally forced to satisfy (on average) a spherical
constraint, $\sum_{i=1}^N s_i^2 = N$, with $N$ the total number of spins~\cite{CrSo92}.
The spherical constraint is imposed on average by adding a term
\begin{equation}
H_{\rm constr} = \frac{z}{2} \ \left(\sum s_i^2 - N \right)
\end{equation}
to the Hamiltonian, with $z$ a Lagrange multiplier. The spins thus defined do not
have an intrinsic  dynamics. In statistical physics applications their temporal evolution
is given by the coupling to a thermal bath, {\it via} a Monte Carlo rule or a
Langevin equation~\cite{CuKu93}.

The model can be endowed with conservative dynamics if one changes the
``spin'' interpretation into a ``particle'' one by adding a kinetic energy~\cite{CuLo98,CuLo99}
\begin{eqnarray}
H_{\rm kin}[\{\dot s_i\}] &=& \frac{m}{2} \sum_{i=1}^N (\dot s_i)^2
\; ,
\label{eq:pspin-kin}
\end{eqnarray}
to the potential energy.
The total energy of the {\it Hamiltonian spherical $p$-spin model}  is then
\begin{equation}
H_{\rm syst} = H_{\rm kin} + H_{\rm pot} + H_{\rm constr}
\; .
\label{eq:pspin-total-energy}
\end{equation}
This model represents a particle constrained to move on an $N$-dimensional hyper-sphere
with radius $\sqrt{N}$. The position of the particle is given by an N-dimensional vector $\vec s=(s_1, \dots, s_N)$ and its velocity
by another $N$-dimensional vector
$\dot{\vec s}=(\dot s_1, \dots, \dot s_N)$.
The $N$ coordinates $s_i$  are globally constrained
to lie, as a vector, on the hypersphere with radius $\sqrt{N}$. The velocity vector $\dot {\vec s}$ is, on average, perpendicular to
$\vec s$, due to the spherical constraint. The parameter $m$ is the mass of the particle.
The parameter $p$ is an integer and we will take it to be equal to $3$ in the numerical applications.

The potential energy (\ref{eq:pspin-pot}) is one instance of a generic random potential $V(\{s_i\})$
with zero mean and correlations~\cite{En93,FrMe94,CuLe96}
\begin{equation}
[ V(\{s_i\})  V(\{s'_i\}) ] = -N {\cal V}(|\vec s-\vec s'|/N)
\end{equation}
with $ {\cal V}(|\vec s-\vec s'|/N) =-  \frac{J^2}{2} (\vec s \cdot \vec s'/N)^p$. The problem is also interesting
for generic ${\cal V}$ but we will focus here on the
monomial case that corresponds to the $p$-spin model. Details on the changes induced by other functions
${\cal V}$ will be given elsewhere.

The generic set of $N$ equations of motion for the system coupled to a white bath is
\begin{equation}
m \ddot s_i(t) + \gamma \dot s_i(t) + z(t) s_i(t) = - \sum_{(i_2 < \dots < i_p)\neq i} J_{i i_2 \dots i_p} s_{i_2}(t) \dots s_{i_p}(t) + \xi_i(t)
\label{eq:dynamic-p2}
\end{equation}
where the random force $\xi_i$ has a Gaussian distribution with zero mean and correlations $\langle \xi_i(t) \xi_j(t') \rangle = 2 \gamma T \delta(t-t')$,
$\gamma$ is the friction coefficient and $T$ is temperature. We have set, and we will keep, the Boltzmann constant to be
$k_B=1$. We have added the friction and
noise terms to the Newton equation for completeness and to make contact with the stochastic setting usually used in
the study of this model. Still,
having introduced in Eq.~(\ref{eq:pspin-kin}) the kinetic energy that provides an intrinsic dynamics to the system,
we will be able to switch off the coupling to the bath, that is to say set $\gamma=0$, and study the dynamics of the isolated system.

The initial condition
will be taken to be $\{s_i^0, {\dot s}_i^0\} \equiv \{ s_i(0), {\dot s}_i(0)\}$ and chosen in ways that we specify below.

\subsection{The $p\geq 3$ case}

The spherical $p$-spin model behaves very differently for $p=2$ and $p\geq 3$. Here we concentrate on the
cases $p\geq 3$ and we leave for a future study the $p=2$ model.

The model with $p\geq 3$ has a very rich and complex free-energy landscape with interesting metastability. In the past, the
model with just the potential energy was analysed in a considerable degree of detail. The kinetic energy allows one to study the dynamics
of the isolated system without changing the picture of metastability described below, since it
contributes a trivial Maxwell (Gaussian) factor to the canonical probability weight.

In short, the known results for the spherical $p\geq 3$ model can be summarised as follows.
In the thermodynamic limit, $N\to\infty$,  the model has two ``critical'' temperatures
$T_s < T_d$. At high temperatures, $T>T_d$,
with  $T_d$ the {\em dynamical} critical temperature,
 the equilibrium state is the paramagnetic one, with vanishing local order parameters.
The analysis of the order-parameter dependent free-energy landscape
proves that there is an exponentially large number of metastable states above
the {\em static} critical temperature
 $T_s$, (and until some high temperature $T_{\rm max}$).
Between $T_d$ and $T_s$ equilibrium is dominated
by a class of metastable states that exist in exponentially large number (finite complexity). Below $T_s$
the number of metastable states is no longer exponential and the Gibbs-Boltzmann measure is
dominated by the low lying ones. The dissipative relaxation dynamics of the model is consistent
with the picture emerging from the static analysis. It is also surprising since for quenches from above to below $T_d$ the
relaxation is non-stationary with ageing effects and other special features due to the fact that it occurs, asymptotically,
on a flat region of phase space.

\begin{figure}[h]
\begin{minipage}[c]{1.00\linewidth}
\begin{center}
\includegraphics[scale=0.8]{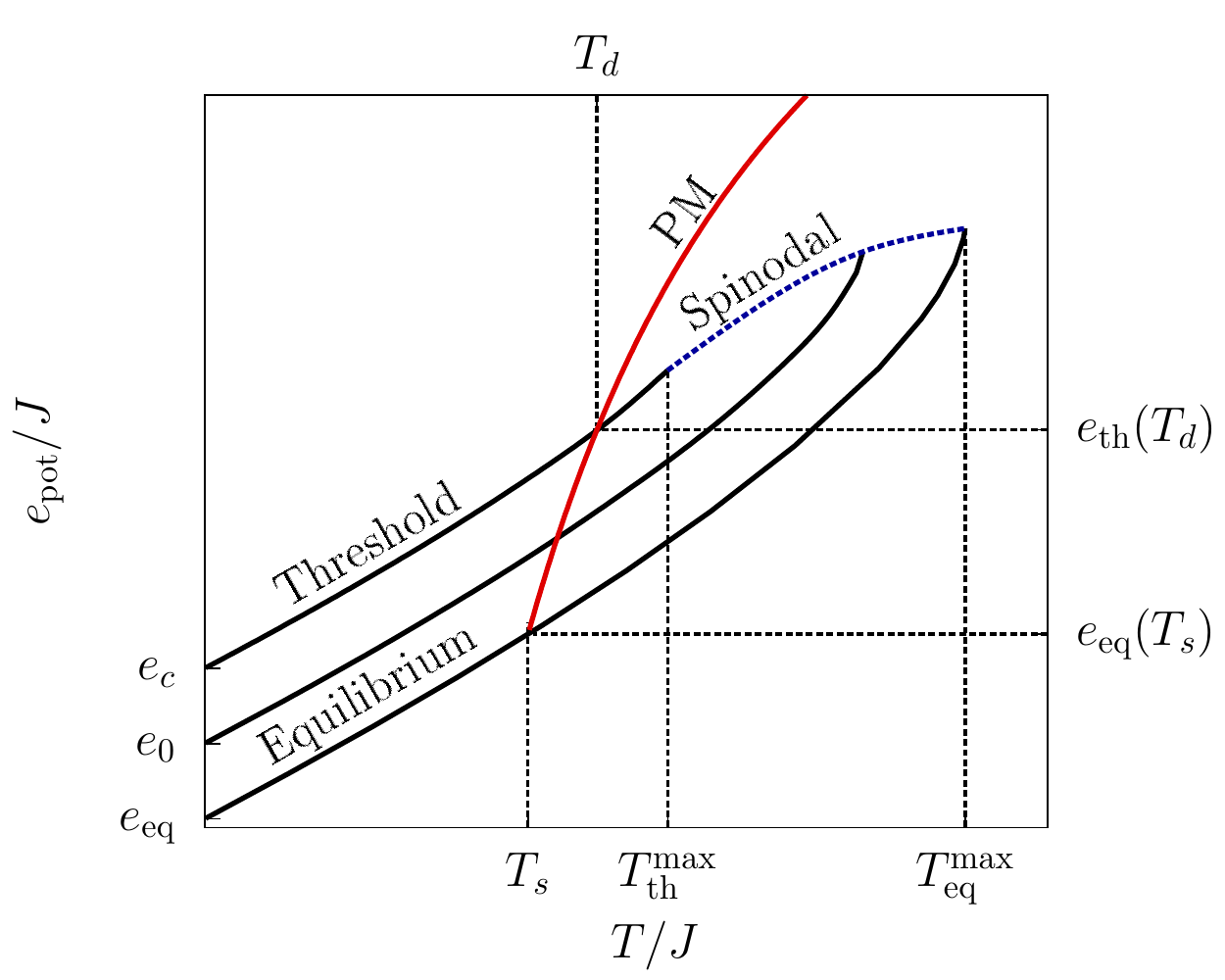}
\end{center}
\end{minipage}

\vspace{0.5cm}

\begin{minipage}[c]{1.00\linewidth}
\begin{center}
\begin{tabular}{|c|c|c|c|c|c|}
\hline
& $T=0$ & $T_s = 0.586$ & $T_d =0.6123$ &  $T_{\rm th}^{\rm max} = 2/3$ & $T_{\rm eq}^{\rm max} = 0.7912$
\\
\hline
\hline
$e_{\rm th}$ &
$-1.155$ &
$-0.853$ &
$-0.839$ &
$-0.778$ &
\\
\hline
$q_{\rm th}$ &
1\;\;\; &
\; 0.539 &
\; 0.500 &
1/3 &
\\
\hline
$e_{\rm eq}$
& $-1.172$
& $-0.853$
& $-0.817$
& $-0.750$
& $-0.632$
\\
\hline
$q_{\rm ea}$
& 1 \;\;\;
& \; 0.645
& \; 0.622
& \; 0.570
& 1/3
\\
\hline
\end{tabular}
\end{center}

\caption{\small Sketch of the potential energy $e_{\rm pot}$
as a function of temperature $T$. The lines
represent the temperature dependence of the potential energy for some selected states (the equilibrium state, the threshold state and an intermediate state).
We also show the paramagnetic energy (with a full red line)
and the spinodal line (with a dotted blue line). We  mark on the plot all the relevant energies and temperatures.
Notice that the potential energy is negative.
In the Table we show some typical values of temperatures and energies in units of $J$ for the
$p=3$ spherical model.
}
\label{fig:sketch-pot-energy}
\label{table:values}
\end{minipage}

\end{figure}

In the rest of this Section we turn the description in the previous paragraph quantitative. We give some
details on how to derive the picture above with three methods: the replica trick, the Thouless-Anderson-Palmer (TAP) approach and
the dissipative dynamics. We give the expressions for some characteristic temperatures and energies that will be useful in
the study of the quench dynamics of the isolated system and we sum up, in Table~\ref{table:values}, the values these take in
the $p=3$ case.

\subsubsection{Equilibrium}
\label{subsec:equilibrium}

The static properties of a model with quenched randomness follow from the study of the disorder-averaged free-energy density
in the canonical ensemble. The kinetic energy term in
(\ref{eq:pspin-kin}) is a conventional one and it does not depend on the quenched disordered interactions.
The statistical sum over the velocities
in the partition function can be readily performed and it yields the usual factor stemming from the Maxwell weight, $[2\pi/(\beta m)]^{N/2}$. Instead,
the contribution of the potential energy to the average over disorder of the logarithm of the partition function
has been computed with
the help of the replica trick~\cite{CrSo92}. In this framework one introduces an $n\times n$ ``overlap''  matrix
$Q_{ab} = N^{-1} \sum_{i=1}^N [\langle s^a_i s^b_i\rangle]$ between replicas $a$ and $b$ ($a,b=1,\dots,n$)
where the angular brackets denote the statistical average. In the case of the $p\geq 3$ model,
the {\it one step replica symmetric Ansatz} solves this problem exactly below $T_s$ in the $N\to\infty$ limit. Within this {\it Ansatz}  the
matrix is parametrized as
$Q_{ab} =
\delta_{ab} + q_{\rm ea} \epsilon_{ab}
$,
with $\epsilon_{ab}$ equal to one if $ab$ are within a diagonal square block of size $m_{\rm eq}\times m_{\rm eq}$ and zero otherwise.
The parameters $m_{\rm eq}$ and $q_{\rm ea}$ are determined by requiring that they extremise the free-energy density
calculated in the limit $n\to 0$ (after $N\to\infty$).

The Edwards-Anderson parameter, $q_{\rm ea}$, and the
replica parameter, $m_{\rm eq}$ were derived in~\cite{CrSo92}. They are given,
in implicit form, by the equations
\begin{eqnarray}
 \frac{1}{p} &=& - y \ \frac{1-y + \ln y}{(1-y)^2}
\; ,
\label{eq:y}
\\
\frac{p}{2} q_{\rm ea}^{p-2} (1-q_{\rm ea})^2 &=& y \ \frac{T^2}{J^2}
\; ,
\label{eq:qeq}
\\
m_{\rm eq} &=& \frac{1-y}{y} \ \frac{1-q_{\rm ea}}{q_{\rm ea}}
\; ,
\label{eq:meq}
\end{eqnarray}
that are easy to solve numerically.
In the $T\to 0$ limit, $q_{\rm ea} = 1-a T$ and $m_{\rm eq} = b T$ with $a=(2y/p)^{1/2}$ and $b=a (1-y)/y $.

The static transition temperature occurs at $T_s$ such that $m_{\rm eq}=1$. This condition implies
\begin{equation}
q_s = q_{\rm ea}(T_s) = 1-y
\end{equation}
that in turn fixes $T_s$ to~\cite{CrSo92,Ba97,CaCa05}
\begin{equation}
T_s  = \sqrt{\frac{p y}{2}}(1-y)^{\frac{p}{2}-1} \ J \; .
\end{equation}
Specializing to the $p=3$ case,
$
T_s = 0.586 \ J
$. The phase transition is discontinuous, in the sense that the order parameter $q_{\rm ea}$ jumps at $T_s$,
$\lim_{T\to T^-_s} q_{\rm ea}(T)\neq 0$ and
$\lim_{T\to T^-_s} m_{\rm eq}(T) =1$ while above $T_s$, $q_{\rm ea}(T) = 0$ and $m_{\rm eq}(T) $ remains fixed to $1$. The
solution boils down, at high temperature, to a replica symmetric one.
However, there are no jumps in the thermodynamic quantities so  the transition  {\it is not} of first order in this sense.
It is a ``random first order phase transition''.

The equilibrium potential energy density, $e^{\rm eq}_{\rm pot} \equiv \langle H_{\rm pot}\rangle$,
is then
\begin{eqnarray}
e^{\rm eq}_{\rm pot} (T) =
- \frac{J^2}{2T} \left[ 1-(1-m_{\rm eq}) q_{\rm ea}^p \right]
\end{eqnarray}
that simplifies to $e^{\rm eq}_{\rm pot} = - J^2/(2T)$ above $T_s$ and takes the value $e^{\rm eq}_{\rm pot}(T_s) = -0.853 \ J $
at $T_s$ for $p=3$.
At zero temperature the equilibrium potential energy is
\begin{equation}
e^{\rm eq}_{\rm pot}(T\to 0) = - \frac{J}{\sqrt{2py}} \; [ 1 + (p-1) y]
\label{eq:equilibrium-e}
\end{equation}
and  $e^{\rm eq}_{\rm pot} = -1.172 \ J $ at $T=0$ for $p=3$.

Above $T_s$ the $q_{\rm ea} \neq 0$ solution still exists (with $m_{\rm eq}>1$) and it disappears at a much higher temperature, $T_{\rm eq}^{\rm max}$,
where
\begin{equation}
q_{\rm ea}^{\rm max} = \frac{p-2}{p}
\qquad\qquad
\mbox{and}
\qquad\qquad
T_{\rm eq}^{\rm max} =  \sqrt{\frac{2}{yp}} \, \left( \frac{p-2}{p} \right)^{(p-2)/2} \ J
\; .
\label{eq:Teqmax}
\end{equation}
For $p=3$, $T_{\rm eq}^{\rm max} = 0.791 \ J$.

\subsubsection{Metastability}
\label{subsubsec:metastability}

The Thouless-Anderson-Palmer (TAP) method serves to derive mean-field equations for the
local order parameters, the local magnetisations,
\begin{equation}
m_i = \langle s_i\rangle
\; ,
\end{equation}
at fixed quenched disorder in the canonical ensemble. As everywhere in the paper, the angular brackets denote statistical average.

The TAP method also constructs the full free-energy and potential energy density landscapes.
It has been applied to the study of the  disordered $p$-spin models landscapes
in, e.g.,~\cite{Ri92,KuPaVi92,CaGiPa98}.
The outcome for the potential energy as a function of temperature is summarised in Fig.~\ref{fig:sketch-pot-energy}.
We explain the meaning of the different lines, and the special values of the potential energy and temperature
highlighted in the figure, in the rest of this Subsubsection.

First of all, it is convenient to introduce
\begin{equation}
q \equiv \frac{1}{N} \sum_{i=1}^N m^2_i
\end{equation}
and then observe that, for the spherical $p$-spin model~\cite{KuPaVi92},
\begin{equation}
H_{\rm pot}[\{m_i\}] = q^{p/2} H_{\rm pot}[\{s_i\}]
\; ,
\end{equation}
due to the homogeneity of the potential energy. This property is not general and makes the structure of the
potential energy landscape of the spherical $p$-spin model particularly simple, with no level crossings nor birth of states at finite temperature,
as sketched in Fig.~\ref{fig:sketch-pot-energy}. In other words, each state can be univocally labeled by its zero-temperature
potential energy.

The extremisation conditions on the TAP free-energy  lead to~\cite{KuPaVi92}
\begin{eqnarray}
(p-1) (1-q) \;  q^{(p-2)/2} = \frac{T}{J^2}  \left( -e_0 - \sqrt{e_0^2 - e_c^2} \right),
\label{eq:tap}
\end{eqnarray}
with
\begin{equation}
e_c = - \sqrt{\frac{2(p-1)}{p}} \ J \; ,
\label{eq:ec}
\end{equation}
and $e_0$ the zero temperature energy.
This equation admits real solutions for $e_0$ such that $e_{\rm eq}(T=0)< e_0 < e_c<0$, with $e_c$ the {\it threshold}
energy at zero temperature. Each of these solutions is an equilibrium or metastable state of the system.
Equation~(\ref{eq:tap}) determines $q$ as a function of the energy density of a TAP solution at $T=0$, $e_0$, and temperature, $T$, both measured in
units of $J$.
In particular, one can check that replacing $e_0$ by $e_{\rm eq}(T=0)$ as given by Eq.~(\ref{eq:equilibrium-e}),
the equation fixing the equilibrium Edwards-Anderson parameter, Eq.~(\ref{eq:qeq}), is recovered.
Otherwise, replacing $e_0$ by $e_c$ the equation for the threshold $q_{\rm th}$ is obtained.
The left-hand-side (l.h.s.) has a bell-shaped form with a maximum at $q_{\rm max}=(p-2)/p$.
For $q=q_{\rm max}$ the same equation determines $T_{\rm max}(e_0)$ above which
the equation has no more solution. The physical solution is the one taking the largest value of $q$,  since it
continuously connects to the zero-temperature one in which $q=1$.
 Different TAP states cease to exist at different temperatures,
as sketched in Fig.~\ref{fig:sketch-pot-energy}. The temperature at which this occurs for the equilibrium level is given in
Eq.~(\ref{eq:Teqmax}). The threshold level, instead, disappears at
\begin{equation}
T^{\rm max}_{\rm th} = \sqrt{\frac{2}{p} (p-1)} \left( \frac{p-2}{p} \right)^{\frac{p-2}{2}} \ J
\label{eq:TmaxTh}
\end{equation}
and $T^{\rm max}_{\rm th} = 2/3 \ J$ for $p=3$.

An expression of the potential energy $e$ as a function of $e_0$ and $q$ (itself a function of $T$) is
\begin{equation}
e(T,e_0) = q^{p/2} e_0 - \frac{J^2}{2T} \left[ (p-1) q^p - p q^{p-1} + 1 \right]
\; .
\end{equation}
Of particular interest is the threshold potential energy that we will give explicitly in Eq.~(\ref{eq:e-th2}).


The equation that fixes the $q$ values of the TAP states that combine (because of their macroscopic degeneracy) to yield
equilibrium in the range $T_s < T < T_d$ is~\cite{KuPaVi92,Ba97}
\begin{equation}
\frac{p}{2} \, q^{p-2} (1-q) = \frac{T^2}{J^2}
\; .
\label{eq:q-eq-no-quench}
\end{equation}
(These states have thermodynamic properties, like the internal energy, that coincide with the continuation of the
high-temperature paramagnetic one. In the sketch in Fig.~\ref{fig:sketch-pot-energy} these states correspond to points along the continuation of the PM line (red) inside the region where metastable states exist, i.e., the portion of the PM curve between $T_s$ and $T_d$.
Once again, the bell-shape form of the  l.h.s. of Eq.~(\ref{eq:q-eq-no-quench}) implies that this equation
admits a solution $q\neq 0$ for temperatures such that $T<T^{\rm max}$ with
\begin{equation}
(T^{\rm max})^2 = a^2(p) J^2
\; .
\label{eq:Tmax-pm}
\end{equation}
were we have introduced the numerical constant
\begin{equation}
a^2(p)= \frac{p}{2}  \frac{(p-2)^{p-2}}{(p-1)^{p-1}}
\label{eq:adep}
\end{equation}
that is smaller than $1$ for all $p\geq 3$.

We will recall the dynamic significance of these TAP states in the next Subsubsection. We announce here that
$T^{\rm max}$ and $q(T^{\rm max})$ coincide with $T_d$ and $q_d$, the critical dynamical temperature and the
value of the parameter $q$ at this temperature.
In particular, for $p=3$, $T^{\rm max} = \sqrt{3/8} \ J \simeq 0.612  \ J = T_d$ and at this temperature, $q=q_{\rm max} = 1/2$.

The scenario with a large multiplicity of metastable states has been confirmed with the exhaustive enumeration of the
extrema of the TAP potential energy landscape of finite (small) size
spherical $p$-spin models at fixed randomness~\cite{meStKa13}.

\subsubsection{Relaxation dynamics}
\label{subsubsec:relaxation-dynamics}

The over-damped relaxation dynamics of the spherical $p$-spin model (coupled to a Markovian bath) were
studied in~\cite{CuKu93,CuKu95}.
The dynamics considered in these papers evolve a completely random initial condition, $\{s_i^0\}$,
that, for purely relaxational dynamics corresponds  (formally)  to an infinite temperature initial
state. The latter is then quenched to a final state in contact  with a bath at finite temperature $T$. The analysis is
performed in the thermodynamic limit, $N\to\infty$, and times are taken to infinity only after, remaining therefore
finite with respect to $N$.

For quenches with   $T>T_d$ the dynamics quickly approach equilibrium
at the new temperature. The correlation and linear response are invariant under translations of time and they are related by the
fluctuation dissipation theorem.

Above but close to $T_d$ the relaxation exhibits a strong slowing down with the correlation
decaying in two steps, with a first approach to a plateau and
a further decay from this plateau, in a much longer time-scale, to zero. This is similar to
what is observed in super-cooled liquids and it is the reason why this model has been
used as a toy model for glass formers of fragile kind.

For $T<T_d$, the evolution of the correlation and linear response functions conform to the {\it weak-egodicity breaking}
scenario~\cite{CuKu93,CuKu95} in which they separate in two contributions evolving in different two-time regimes
\begin{eqnarray}
C(t_1,t_2) &=& C_{\rm st} (t_1 - t_2) + C_{\rm ag}(t_1,t_2)
\; ,
\label{eq:C-sep}
\\
R(t_1,t_2) &=& R_{\rm st} (t_1 - t_2) + R_{\rm ag}(t_1,t_2)
\; ,
\label{eq:R-sep}
\end{eqnarray}
with the stationary and a non-stationary terms linked by the fluctuation-dissipation theorem (FDT) at the temperature of the
bath and a modified FDT at an effective temperature $T_{\rm eff}$~\cite{CuKuPe97,Cu11} selected by the dynamics,
\begin{eqnarray}
R_{\rm st}(t_1-t_2) &=& - \frac{1}{T} \frac{d C_{\rm st}(t_1-t_2)}{d(t_1-t_2)}
\; , \\
R_{\rm ag}(t_1-t_2) &=& \frac{1}{T_{\rm eff}} \frac{\partial C_{\rm ag}(t_1,t_2)}{\partial t_2}
\; ,
\end{eqnarray}
always with $t_1\geq t_2$. In the asymptotic limit, the two terms added to form $C$ and $R$ evolve in different regimes in the
sense that when one changes the other one is constant and vice versa.
The limiting values of the various contributions to the correlation function
are
\begin{eqnarray}
C_{\rm st}(0) = 1-q \; , \qquad\qquad &
\lim_{\tau \to \infty} C_{\rm st}(\tau) = 0
\; ,
\label{eq:lim-stat}
\\
\lim_{t_2\to t_1^-} C_{\rm ag}(t_1,t_2) = q \; , \qquad\qquad &
\lim_{t_1\gg t_2} C_{\rm ag}(t_1,t_2) = 0
\; ,
\label{eq:lim-aging}
\end{eqnarray}
with $q$  being equal to $q_{\rm th}$,
the value at the threshold of the TAP free-energy density.

The asymptotic potential energy reached after the quench for $T<T_d$ is
the one of the threshold level in the free-energy landscape:
\begin{equation}
e_{\rm th}
= - \frac{J^2}{2} \left[ \frac{1}{T} (1-q_{\rm th}^p) + \frac{1}{T_{\rm eff}} q_{\rm th}^p \right]
\; .
\label{eq:e-th2}
\end{equation}
 This expression will be very useful in the numerical analysis.
The parameters $q_{\rm th}$ and $T_{\rm eff}$ are given by
\begin{eqnarray}
\frac{p (p-1)}{2} (1-q_{\rm th})^2 q_{\rm th}^{p-2} &=& \frac{T^2}{J^2}
\; ,
\label{eq:qth}
\\
m_{\rm th} \; \equiv \; \frac{T}{T_{\rm eff}} &=&  \frac{(p-2)(1-q_{\rm th})}{q_{\rm th}}
\; .
\label{eq:mth}
\end{eqnarray}
The parameter $m_{\rm th}$ measures the violation of the fluctuation-dissipation theorem out of equilibrium and can be interpreted in
terms of an effective temperature $T_{\rm eff}$~\cite{CuKuPe97}  (note that $m_{\rm th} \leq 1$ and
$T_{\rm eff} > T$ for quenches from high to low temperature).

The bell-shaped function in the l.h.s. of Eq.~(\ref{eq:qth}) indicates, once again, that this equation has two solutions until a temperature
$T^{\rm max}_{\rm th}$ given by the same expression, Eq.~(\ref{eq:TmaxTh}),  obtained with the TAP formalism. The relevant solution is the
one taking the higher value, the one that is connected to $q_{\rm th}=1$ at $T=0$.
Its value and the energy at this temperature are given in the Table that accompanies Fig.~\ref{table:values}.

The equation fixing $q_{\rm th}$ as a function of $T$ is implicit so we cannot write an explicit expression for
$e_{\rm th}(T)$. We can, instead, eliminate $T$ from Eq.~(\ref{eq:e-th2}) using Eqs.~(\ref{eq:qth}) and (\ref{eq:mth}), and then
recast Eq.~(\ref{eq:e-th2})  as
\begin{equation}
e_{\rm th} = - \frac{J}{\sqrt{2p(p-1)}} \ q_{\rm th}^{(2-p)/2} \left[ \frac{1-q_{\rm th}^p}{1-q_{\rm th}} + (p-2) q_{\rm th}^{p-1} \right]
\; .
\label{eq:e-th-recast}
\end{equation}
At $T\simeq 0$, $q_{\rm th}\simeq 1- aT$ with $a^2=2/(p (p-1))$ and $m_{\rm th} \simeq (p-2) aT$.
Expanding in powers of $T$ one finds
$e_{\rm th} \simeq - a (p-1)/2$ and then
\begin{equation}
e_{\rm th}(T=0)=- \sqrt{\frac{2(p-1)}{p}} \ J \equiv e_c
\end{equation}
as expected.
Its concrete value for $p=3$ is given in Table~\ref{table:values}.

The dynamic critical temperature arises when $m_{\rm th}=1$:
\begin{equation}
T_d = a(p)\ J
\label{eq:Td}
\end{equation}
that coincides with Eq.~(\ref{eq:Tmax-pm}).
Specialising to the $p=3$ case
$
T_d = 0.612 \ J
$.
When this occurs
$e_{\rm th}(T_d)=e_{\rm eq}(T_d)=-J^2/(2T_d)$ and $q=q_d=q_{\rm th}(T_d)=(p-2)/(p-1)$.

The dynamic relevance of the TAP states that are non trivial but ``confused'' with the paramagnetic solution with the conventional replica calculation,
the ones with  $q$ values determined by Eq.~(\ref{eq:q-eq-no-quench}), is understood from the analysis of the relaxation dynamics starting from initial conditions
in equilibrium at the range of temperature $T_s< T'< T_d$~\cite{HoJaYo83,ThKi88,FrPa95,BaBuMe96}.
These initial conditions are confined within TAP states that do not let the system
escape. The asymptotic dynamics remain within the departing state, as indicated by the fact that $\lim_{t_1\to\infty} \lim_{t_2\to\infty} C(t_1,t_2) =q$, with $q$ given by
Eq.~(\ref{eq:q-eq-no-quench}).

For finite $N$ the threshold level dynamics and evolution within TAP states should have finite, though exponentially large in $N$, lifetime. Some numerical
evidence for this was given in, e.g.,~\cite{CuKuLePe97,BeCuIg01,Be03}.

\section{Quenches and dynamics of the isolated system}
\label{sec:quenches}

The Hamiltonian (\ref{eq:pspin-total-energy}) with $H_{\rm kin}$ and $H_{\rm pot}$ given in Eqs.~(\ref{eq:pspin-kin}) and (\ref{eq:pspin-pot})
has two parameters, the variance of the couplings $J_{i_1\dots i_p}$ and the mass of the particle. We will
consider initial conditions sampled from equilibrium at $T'$ with Hamiltonian $H_0$ and evolve them with a different Hamiltonian
$H$. Since the potential and kinetic energies play different roles, as the former depends on the quenched randomness
while the latter does not, the treatment of the quenches induced by a change in the random interactions needs a bit more
care.

Let us take the $p$-spin model  (\ref{eq:pspin-pot}) and (\ref{eq:pspin-kin})
with coupling constants $J^0_{i_1 \dots i_p}$ in canonical equilibrium at temperature $T'$ and
evolve it in isolation from the environment with a modified Hamiltonian. Quenches in the random exchanges that do not
keep any memory of the values before the quench are not interesting.
We therefore impose quenches in the potential energy
such that each random choice of the exchanges is changed into $J_{i_1 \dots i_p}$, with the new couplings related to the old ones by
\begin{equation}
J^0_{i_1 \dots i_p} =  J_{i_1 \dots i_p} \ x \qquad\qquad \forall \; {i_1,...,i_p} \; .
\end{equation}
This transformation is such that for each sample (disorder
realisation) at $t=0$ we uniformly change the value of all random
couplings by the same factor. In other words, we prepare the system in
a thermal state of a Hamiltonian with potential energy
\[
H^{0}_{\rm pot}[\{s_i\}]=-\sum_{1\le i_{1}<\ldots<i_{p}\le N}J{}_{i_{1}\ldots i_{p}}^{0}s_{i_{1}}\ldots s_{i_{p}}
\; ,
\]
but let each initial condition sampled from this state evolve with the Hamiltonian with potential energy
\[
H_{\rm pot}[\{s_i\}]=-\sum_{1\le i_{1}<\ldots<i_{p}\le N}J{}_{i_{1}\ldots i_{p}}s_{i_{1}}\ldots s_{i_{p}}
\; .
\]
 Note that with $x<1$ we  enhance the interactions and with $x>1$
we depress the interactions between the spins.
Technically, after relating the coupling strengths one by one we  have only one quenched disorder average to make.

It is important to note that under this change, the  potential energy levels in Fig.~\ref{fig:sketch-pot-energy} are translated upwards or downwards,
and stretched or contracted, for $x<1$ or $x>1$, respectively. Indeed, one can easily see the translation by noticing
that $e_0$ is proportional to $J$, and the contraction by noticing that the various $T_{\rm max}$ are proportional to $J$. Concomitantly,
the static and dynamic critical temperatures $T^0_s=T_s(J_0)$ and $T^0_d=T_d(J_0)$ of the initial potential
are shifted to new values $T_s=T_s(J)$ and $T_d=T_d(J)$ after the quench,
\begin{equation}
\frac{T_s}{T_s^0} = \frac{J}{J_0} = \frac{1}{x} \qquad \Rightarrow \qquad
T_s^0 = x \ T_s
\qquad\mbox{and}\qquad
T_d^0 = x \ T_d
\; .
\end{equation}

We will also consider quenches in the mass, $m_0 \mapsto m$, that change the kinetic contribution to the energy as
\begin{equation}
H^0_{\rm kin} = \frac{m_0}{2} \sum_i {{\dot s}_i}^2
\qquad\mapsto\qquad
H_{\rm kin} = \frac{m}{2} \sum_i {{\dot s}_i}^2
\end{equation}

\subsection{Dynamical equations}

Importantly enough, all our results will be derived after having taken the limit $N\to\infty$
from the start, and eventually considering the long-times asymptotic limit only after.

In the limit $N\to\infty$ the dynamics of the model are fully characterised by the behaviour of the
two-time correlation and linear response function. The equations ruling their evolution are  easily
derived using the Martin-Siggia-Rose functional formalism.

Particular initial conditions can be imposed by including in the dynamic generating function an
integration over the initial conditions weighted with their distribution. Equilibrium initial conditions
at a temperature $T' = {\beta'}^{-1}$ are distributed according to the Gibbs-Boltzmann measure
 \begin{equation}
P(\{s_i(0), \dot s_i(0) \}) = Z^{-1}(\beta') \, {\rm e}^{-\beta' H_{0}(\{s_i(0),\dot s_i(0))\}}
 \end{equation}
 with $H_0$ defined in Eqs.~(\ref{eq:pspin-pot})-(\ref{eq:pspin-total-energy}). The
 Hamiltonian depends on the quenched random interactions. The average over disorder
in the case of initial states correlated with the quenched randomness needs the use of the replica trick, as
 explained in Ref.~\cite{HoJaYo83}. This means that the spin variables evaluated at the initial time have to be replicated,
 $s_i(0) \mapsto s_i^a(0)$, with $a=1,\dots, n$, to perform the average over the random exchanges.
 The subsequent evolution of each of these replicas has to be followed in time, and it turns out that
 the replica structure of the initial condition is conserved.

 For the dissipative spherical $p$-spin model this calculation has been carried out in~\cite{FrPa95,BaBuMe96},
 and it can be adapted to the isolated model with kinetic energy with just minor modifications.
 We therefore present the outcome here without giving many details of the derivation. In order to facilitate the comparison to the
 expressions for the dissipative model  we keep the coupling to the white
bath active in the presentation of the dynamic equations.
Later on, we will focus on the isolated problem.

In the $N\to\infty$ limit, the only relevant correlation and linear response functions that determine the dynamics  of the
model are
\begin{eqnarray}
\qquad\qquad
C_{ab}(t_1,t_2) &=&  N^{-1} \sum_{i=1}^N \; [\langle s^a_i(t_1) s^b_i(t_2)\rangle] \; ,
\\
\qquad\qquad
C_{ab}(t_1,0) &=&  N^{-1} \sum_{i=1}^N \; [\langle s^a_i(t_1) s^b_i(0)\rangle] \; ,
\\
\qquad\qquad
R_{ab}(t_1,t_2) &=&  \left. N^{-1} \frac{\delta \;\;\;}{\delta h_b(t_2)} \sum_{i=1}^N \; [\langle {s^a}^{(h)}_{\!\!\! i}(t_1)\rangle]
\right|_{h=0} \; ,
\end{eqnarray}
for $t_1, \ t_2 > 0$,
where the infinitesimal perturbation $h$ is coupled linearly to the spin $H\mapsto H - h \sum_{i=1}^N s_i$ at time $t_2$
and the upperscript ${(h)}$ indicates that the configuration is measured after having applied the field $h$.
The square brackets denote here and everywhere in the paper the average over quenched disorder. The angular brackets indicate the average over
thermal noise if the system is coupled to an environment, and over the initial conditions of the dynamics sampled with the
probability distribution $P$. When the coupling to the bath is set to zero, $\gamma=0$, the last average is the only one remaining in the
angular brackets operation.

Without loss of generality we will focus on initial states in equilibrium at $T' \geq T^0_s$, where the replica structure is symmetric, although there
can still be a complex structure of metastable states, as explained in Sec.~\ref{subsubsec:metastability}. Considering the case $T'<T^0_s$ would add quite a lot of
unnecessary complexity to the calculations, while we do not expect major changes in the dynamic behaviour
under such initial conditions. For these reasons, the following expressions are
valid only for  $T'\geq T^0_s$.

The dynamical equations of the model coupled to a bath at temperature $T$, starting from a random state,  are
well known and can be found in
Refs.~\cite{CuKu93,Ba97,CaCa05,CuKu95}. They can be derived from the
dynamical Martin-Siggia-Rose action
\begin{equation}
S_{J}=\sum^{N}_{i=1}\int^{\infty}_{0} \!\!\! dt\,\hat{s}_i(t)\left( \gamma T\hat{s}_i(t)+\gamma \dot{s}_{i}(t)+m \ddot{s}_{i}(t)+\frac{\partial H_{\rm pot}}{\partial s_i}+z(t)s_i(t)  \right) \; ,
\label{eq:Tinf-action}
\end{equation}
where the subscript $J$ in $S_{J}$ indicates that the action depends explicitly on the disordered couplings and the
$\hat s_i$ are (imaginary) auxiliary variables used to rewrite a delta function that enforces the validity of the equation of motion in the
path integral. (This form corresponds to the Ito convention in which there is no Jacobian contribution.) We included here the kinetic energy contribution
not present in these publications.

The replicated dynamical action that includes the contribution from the distribution of the initial conditions reads
\begin{eqnarray}
\!\!\!\!\
S_{J}=\sum_{a=1}^n \sum^{N}_{i=1}
\int^{\infty}_{0}\!\!
dt\,\hat{s}^a_i(t)\left( \gamma T\hat{s}^{a}_i(t)+\gamma \dot{s}^{a}_{i}(t)+m \ddot{s}^{a}_{i}(t)+\frac{\partial H_{\rm pot}}{\partial s^{a}_i}+z(t)s^{a}_i(t)  \right)
-\frac{1}{T'}H_{0}[s^{a}_i(0), {\dot s}^a_i(0)]
\; .
\label{eq:dyn_action}
\end{eqnarray}

The kinetic energy term in the initial distribution does not depend on the quenched random interactions, it does not
affect the dynamic  equations, but will appear only in the energetic considerations that we will develop below.

We will now show how to perform the average over the couplings in some detail. Two terms in (\ref{eq:dyn_action}) depend on the
interactions, they are the ones in which $H_{\rm pot}$ appears in the force in the evolution equation and
in the initial contribution $H_0$. We collect them in $S_{\rm dis}$. (Note that we use, as a working
assumption, that $z(t)$ does not depend on $J_{ijk}$.)
The average over disorder of the exponentials of these two terms is
\begin{multline}
[{\rm e}^{S_{\rm dis}}]=
\prod_{i<k<l}\int dJ_{ikl}\;
\exp\left\{-\frac{J^2_{ikl}2N^{p-1}}{2p!J^2}
-\frac{1}{T'}J^{0}_{ikl} \sum_{a}s^{a}_{i}(0)s^{a}_{k}(0)s^{a}_{l}(0)
\right.
\\
\left.
-J_{ikl}\sum_{a}\int^{\infty}_{0} \!\!\!
dt\,(\hat{s}^{a}_{i}(t)s^{a}_{k}(t)s^{a}_{l}(t)+s^{a}_{i}(t)\hat{s}^{a}_{k}(t)s^{a}_{l}(t)+s^{a}_{i}(t)s^{a}_{k}(t)\hat{s}^{a}_{l}(t))
\right\},
\end{multline}
where we have included the Gaussian distribution over the couplings (we average over the final couplings $J_{ikl}$ but the same results
would be obtained had we chosen to average over the initial
ones, $J^{0}_{ikl}$).  We have symmetrised the term originating in $\partial H_{\rm pot}/\partial s^{a}_{i}$.
Following our choice of quench we set $J^{0}_{ikl}=x J_{ikl}\,\forall \{ i,k,l \}$, in which case the disorder dependent part of the action becomes
\begin{multline}
 [{\rm e}^{S_{\rm dis}}] \; =\prod_{i<k<l}\int dJ_{ikl}\;
 \exp\left\{-\frac{J^2_{ikl}2N^{p-1}}{2p!J^2}
-
\frac{x}{T'}J_{ikl} \sum_{a}s^{a}_{i}(0)s^{a}_{k}(0)s^{a}_{l}(0)
\right.
\\
\left.
-J_{ikl}\sum_{a}\int^{\infty}_{0} \!\!\! dt\,(\hat{s}^{a}_{i}(t)s^{a}_{k}(t)s^{a}_{l}(t)+s^{a}_{i}(t)\hat{s}^{a}_{k}(t)s^{a}_{l}(t)+s^{a}_{i}(t)s^{a}_{k}(t)\hat{s}^{a}_{l}(t))
\right\}
\; .
\end{multline}
After performing the Gaussian integration over the couplings we have
\begin{eqnarray}
&
\displaystyle{ [{\rm e}^{S_{\rm dis}}]
\; \propto
\prod_{i<k<l}
\;
\exp\left\{\frac{p!J^2}{4N^{p-1}}
\!
\left[
\int^{\infty}_{0} \!\!\! dt\,\sum_{a}(\hat{s}^{a}_{i}(t)s^{a}_{k}(t)s^{a}_{l}(t)+s^{a}_{i}(t)\hat{s}^{a}_{k}(t)s^{a}_{l}(t)+s^{a}_{i}(t)s^{a}_{k}(t)\hat{s}^{a}_{l}(t))
\right.
\right.
}
\nonumber\\
&
\displaystyle{
\left.
+\frac{x}{T'}\sum_{a}s^{a}_{i}(0)s^{a}_{k}(0)s^{a}_{l}(0)
\right]
}
\nonumber\\
&
\qquad\qquad\qquad\qquad\qquad\qquad\qquad
\displaystyle{
\times
\left[
\int^{\infty}_{0} \!\!\! dt'\,\sum_{b}(\hat{s}^{b}_{i}(t')s^{b}_{k}(t')s^{b}_{l}(t')+s^{b}_{i}(t')\hat{s}^{b}_{k}(t')s^{b}_{l}(t')+s^{b}_{i}(t')s^{b}_{k}(t')\hat{s}^{b}_{l}(t'))
\right.
}
\nonumber\\
&
\displaystyle{
\left.
\left.
+\frac{x}{T'}\sum_{b}s^{b}_{i}(0)s^{b}_{k}(0)s^{b}_{l}(0) \right]\right\}.
}
\end{eqnarray}

The product between the two terms involving integrals in time produces the terms already present in the dynamical equations starting from a random initial condition.
These terms are proportional to $J^2$. The product between a term with one time integral and one term with the initial condition generates the new terms.
They are proportional to $J^{2} x$,
that is to say $JJ_0$, if we call $J_0=xJ$. The product of the terms involving only the initial conditions yields the equilibrium equations
of the model decoupled from the dynamics, and are proportional to $J_0^2$, as they should.

Taking now $N\to\infty$ one derives the dynamical equations that read
\begin{eqnarray}
\left(m\partial_{t_{1}}^{2}+\gamma\partial_{t_{1}}+z(t_{1})\right)R(t_{1},t_{2}) & = &
\frac{J^2 p(p-1)}{2}\int_{t_{2}}^{t_{1}}dt'R(t_{1},t')C^{p-2}(t_{1},t')R(t',t_{2})
+\delta(t_1-t_2)
\; ,
\label{eq:dyn-eqs-C}\\
\left(m\partial_{t_{1}}^{2}+\gamma\partial_{t_{1}}+z(t_{1})\right)C(t_{1},t_{2}) & = &
\frac{J^2 p(p-1)}{2}\int_{0}^{t_{1}}dt'R(t_{1},t')C^{p-2}(t_{1},t')C(t',t_{2})
+\gamma T R(t_2, t_1)
\nonumber\\
 &  &
+ \frac{J^2 p}{2}\int_{0}^{t_{2}}dt'R(t_{2},t')C^{p-1}(t_{1},t')
+\frac{JJ_0 p}{2T'}C^{p-1}(t_{1},0)C(t_{2},0) \; ,
\;
\label{eq:dyn-eqs-R}\\
z(t_{1}) & = &
-m\partial_{t_{1}}^{2}C(t_{1},t_{2})\vert_{t_{2}\rightarrow t_{1}^{-}}
+ \gamma T \, R(t_1,t_2) \vert_{t_{2}\rightarrow t_{1}^{-}}
\nonumber\\
 &  &
 +
\frac{J^2 p^{2}}{2}\int_{0}^{t_{1}}dt'R(t_{1},t')C^{p-1}(t_{1},t')+
\frac{JJ_0 p}{2T'}C^{p}(t_{1},0)
\label{eq:dyn-eqs-z}
\; .
\end{eqnarray}
One can check that these equations coincide with the ones in \cite{FrPa95,BaBuMe96} when inertia is neglected and $J=J_0$.

In the rest of the paper we switch off the connection to the environment by setting $\gamma=0$.
With inertia and no coupled bath, the equal-time conditions are
\begin{eqnarray*}
C(t_1,t_1) & = & 1 \; ,\\
R(t_1,t_1) & = & 0 \; ,\\
\partial_{t_1}C(t_1,t_2)\vert_{t_2\rightarrow t_1^{-}}=\partial_{t_1}C(t_1,t_2)\vert_{t_2\rightarrow t_1^{+}} & = & 0 \; ,\\
\partial_{t_1}R(t_1,t_2)\vert_{t_2\rightarrow t_1^{-}} & = & \frac{1}{m} \; ,\\
R(t_1,t_2)\vert_{t_2\rightarrow t_1^{+}} & = & 0 \; ,
\end{eqnarray*}
for all times $t_1, t_2$ larger than $0^+$, when the dynamics start.

\subsection{The Lagrange multiplier}

We found convenient to numerically integrate the integro-differential equations that determine the time-evolution
of the system to use an expression of the Lagrange multiplier that trades the second-time
derivative of the correlation function into the total conserved energy after the quench. More precisely,
we proceeded as explained below.

Firstly, we provide an expression for the kinetic energy density,
\begin{equation}
\label{eq:kin}e_{\rm kin}(t_1) \equiv \frac{E_{\rm kin}(t_1)}{N} = \frac{m}{2N}\sum_i\langle {\dot{s}}^2_i(t_1)\rangle.
\end{equation}
Using the definition of $C(t_1,t_2)$, and the fact that, for sufficiently short time differences $(t_1-t_2)$ it is always possible to write
$C(t_1,t_2)= 1-a(t_1-t_2)^2+\mathcal{O}((t_1-t_2)^4)$, with $a>0$, we find that
\begin{equation}
e_{\rm kin}(t_1)=-\frac{m}{2}\partial_{t_{1}}^{2}C(t_{1},t_{2})\vert_{t_{2}\rightarrow t_{1}^{-}}.
\end{equation}

On the other hand, the potential energy is linked to the kinetic energy and
the Lagrange multiplier {\it via} a general equation proven as follows. Take the
microscopic evolution of $s_i(t_1)$, multiply it by $s_i(t_2)$, and take the average over initial conditions:
\begin{equation}
m \ \langle \ddot s_i(t_1) s_i (t_2) \rangle  =  - p \sum_{i_2 \dots i_p} J_{i i_2 \dots i_p} \langle s_i(t_2) s_{i_2}(t_1) \dots s_{i_p}(t_1) \rangle
- z(t_1) \langle s_i(t_1) s_i(t_2)\rangle
\; .
\end{equation}
Summing now over $i$, normalising by $N$, and taking the limit $t_2\to t_1^-$,
\begin{equation}
\frac{m}{N} \lim_{t_2\to t_1^-} \sum_i \langle \ddot s_i(t_1) s_i (t_2) \rangle
=
m \lim_{t_2\to t_1^-} \partial_{t_1^2} C(t_1,t_2)
=  - p e_{\rm pot}(t_1) - z(t_1)
\; .
\label{eq:identity2}
\end{equation}
This implies that
\begin{equation}
e_{\rm pot}(t_1) \equiv \frac{E_{\rm pot}(t_1)}{N} = \frac{ -m\partial_{t_{1}}^{2}C(t_{1},t_{2})\vert_{t_{2}\rightarrow t_{1}^{-}}-z(t_{1})}{p}.
\label{eq:pot}
\end{equation}
The two contributions added together yield the total energy density of the system
\begin{equation}
e(t_1) =e_{\rm kin}(t_1) +e _{\rm pot}(t_1)
\; ,
\label{eq:ef-conservation}
\end{equation}
conserved after the quench.

Rearranging now the equation for $e(t_1)$, Eq.~(\ref{eq:ef-conservation}), with the help of Eq.~(\ref{eq:identity2}),
we obtain a new expression for the Lagrange multiplier
\[
z(t_{1})=-m\left(\frac{p}{2}+1\right)\partial_{t_{1}}^{2}C(t_{1},t_{2})\vert_{t_{2}\rightarrow t_{1}^{-}}-p e(t_1)
\; .
\]
Using now the original equation for $z(t_1)$, Eq. (\ref{eq:dyn-eqs-z}), we can eliminate
the second time  derivative to obtain
\begin{equation}
z(t_{1})=2e(t_1)+  \frac{ J^2 p \left(p+2\right)}{2} \int_{0}^{t_{1}}dt' \; R(t_{1},t')C^{p-1}(t_{1},t')+ \frac{JJ_0 (p+2)}{2T'} \ C^{p}(t_{1},0)
\; .
\label{eq:z-Nicolas}
\end{equation}
It seems that we have simply traded $z(t_1)$ by $e(t_1)$.
However, for an isolated system $e(t_1)=e_f$, a constant.
Then, the last expression allows a straightforward numerical solution of the evolution equations for the
isolated system since it does not involve the second time derivative of the correlation function. In practice, in the
numerical algorithm we fix the total energy $e_f$ and we then integrate the set of coupled integro-differential
equations with a standard Runge-Kutta method. We only have to define which is the total energy density of the system, the subject of the next
two subsections.

\subsection{Energy change}

We now determine  the energy changes induced by a quench in the disorder exchanges
and a quench in the mass of the particle. As these two act separately on the potential and kinetic contributions to the total energy,
the total energy change is the sum of the two variations.

\subsubsection{The energy change after a potential energy quench}

Let us investigate what is the change in energy density induced by the
change in potential energy $J^0_{i_1 \dots i_p} \mapsto J_{i_1\dots i_p}$,
while keeping the mass constant $m_0=m$.

The energy density just before the quench is the energy density of a canonical equilibrium
paramagnetic state at temperature $T'$ and it is given by
\begin{equation}
\label{eq:dyn_en}
e_{i} = e(0^-)=
\frac{T'}{2}-\frac{J_0^2}{2T'}
\; .
\end{equation}
The first term is the equipartition of the kinetic energy
and the second one is the potential energy of the paramagnet in equilibrium. Note that this is still true if we choose $T^0_s<T'<T^0_d$, since, although metastable
states still dominate the energy landscape in that range of temperatures, the thermodynamics of the equilibrium states is indistinguishable from the one of the
paramagnet (see Section~\ref{subsubsec:metastability}).

The energy density at time $t_1=0^+$ right after the instantaneous quench is
\[
e(0^+)=e_{\rm kin}(0^+)+e_{\rm pot}(0^+)=
-\frac{m}{2}\partial_{t_{1}}^{2}C(t_{1},t_{2})\vert_{t_{2}\rightarrow t_{1}^{-}=0^{+}}
+\frac{-m\partial_{t_{1}}^{2}C(t_{1},t_{2})\vert_{t_{2}\rightarrow t_{1}^{-}=0^{+}}-z(0^+)}{p}
\; .
\]
Using the fact that with no mass change the kinetic energy does not vary between $t=0^-$ and $t=0^+$
\[
-\frac{m}{2}\partial_{t_{1}}^{2}C(t_{1},t_{2})\vert_{t_{2}\rightarrow t_{1}^{-}=0^{+}}=\frac{T'}{2}
\; ,
\]
as confirmed numerically in Sec.~\ref{sec:numerical},
and the value of the Lagrange multiplier evaluated from Eq.~(\ref{eq:dyn-eqs-z}) at $t=0^+$
is
\[
z(0^+)=T'+\frac{JJ_0p}{2 T'}
\]
we find
\begin{equation}
e_f \equiv e(0^+) =\frac{T'}{2}-\frac{JJ_0}{2T'}
\; .
\label{eq:final-energy}
\end{equation}

Equations~(\ref{eq:dyn_en}) and (\ref{eq:final-energy}) imply that the amount of energy injected during the instantaneous
quench $J_{i_1\dots i_p}^0 \mapsto J_{i_1\dots i_p}$ is
\[
\Delta e = e_{f}-e_{i} = \Delta e_{\rm pot} = \frac{J_0(J_0-J)}{2T'}
\; .
\]
Therefore, $\Delta e>0$ if $J_0>J$ and $\Delta e<0$ if $J_0<J$.


\subsubsection{The energy change after a quench in the mass}

If we apply a  quench in the mass, $m_0 \mapsto m$,  while leaving the random exchanges fixed, the energy balance is modified.

Imagine that we initialise the system in a paramagnetic or TAP state such that
$e_{\rm pot}(0^-) = -J^2/(2T')$. If we change the mass according to $m_0 \to m$,
the potential energy does not change during the instantaneous quench. Instead,
the kinetic energy does. Right before the quench the kinetic energy density is
\begin{equation}
e_{\rm kin}(0^-) = \frac{m_0}{2} (\dot s_i(0^-))^2 = \frac{T'}{2}
\; ,
\end{equation}
while right after the quench the velocities have not changed but the
mass of the particle has. Therefore,
\begin{equation}
e_{\rm kin}(0^+) = \frac{m}{2} (\dot s_i(0^+))^2= \frac{m}{2} (\dot s_i(0^-))^2 = \frac{m}{m_0} \frac{T'}{2}
\; .
\end{equation}
The total energy after the quench is
\begin{equation}
e_f = \frac{m}{m_0} \frac{T'}{2} - \frac{J^2}{2T'}
\end{equation}
and the energy input by the quench reads
\begin{equation}
\Delta e = e_f - e_i = \Delta e_{\rm kin} = \left(\frac{m}{m_0} -1\right) \frac{T'}{2}
\; .
\end{equation}

Adding together the energy variation due to the the potential and mass quenches, the total energy change
becomes
\begin{equation}
\Delta e_{\rm tot} = \left( \frac{m}{m_0} -1 \right) \frac{T'}{2} + \frac{J_0^2}{2T'} \left(1 - \frac{J}{J_0}\right)
\; .
\end{equation}

\section{Asymptotic analysis}
\label{sec:asymptotic}

Depending on the pre and post quench parameters the system reaches
different asymptotic dynamics. In some cases,  the system reaches a stationary regime but,  for  parameters
carefully tuned, a final state with non stationary   {\em ageing}  behaviour can also be attained.

Before entering into the deduction of the asymptotic equations, we present the general reasoning that we use to find them.

We will first analyse in Sec.~\ref{subsec:stationary}
the cases in which a stationary regime is reached after the quench. This means that \\
\indent
-  We assume time-translational invariance (TTI)
 $C(t_1, t_2) \mapsto C_{\rm st}(\tau)$, $R(t_1, t_2) \mapsto R_{\rm st}(\tau)$ with $\tau=t_1-t_2$ and
\\
\indent
- the fluctuation-dissipation theorem  $R_{\rm st}(\tau) = -1/T_f \ d_\tau C_{\rm st}(\tau)$ for $\tau \geq 0$.
\\
\indent
- We define the asymptotic limits of the correlation with the initial configuration
$\lim_{t_1\to\infty} C(t_1,0)\mapsto q_0$,
\\
\indent
- and between two dynamic configurations $\lim_{\tau\to\infty} C_{\rm st}(\tau) =q$.\\
\indent
- We assume that the kinetic energy density   approaches $\lim_{t_1\to \infty} e_{\rm kin}(t_1) = T_f/2$ after the quench.
\\
Clearly, all these assumptions can and have been verified numerically.
The temperature of the final state, $T_f$, has to be calculated and the parameters $q_0$ and $q$ as well.

We anticipate that the parameters $q_0$ and $q$ will find two interesting interpretations in the cases
in which the system is initially in a non-trivial TAP state.
The value $q_0$ represents the overlap between a typical configuration of the
TAP state of the pre-quench potential in which the system was prepared initially,
and a typical  configuration of the TAP state into which the original state has evolved
in the post quench potential, if it still exists. Instead, $q$ is the self-overlap within the TAP
state of the post-quench potential. This description will become clear after presenting the analytical and
numerical results.

We will then analyse, in Sec.~\ref{subsec:nonstationary}, the cases in which the system,
starting from equilibrium in a disordered paramagnetic state at high temperature
 is set, after the  quench, on the threshold level and the
stationarity assumption fails. This is in agreement with what was expected from the properties of the states on the threshold, that
are flat, and on which ageing properties were obtained after quenches from random initial conditions in the dissipative setting~\cite{CuKu93,CuKu95}.
For these cases we need to modify the assumptions above and allow for a two-time scale dependence of the correlation and
linear response functions that take a
form with a separation of time-scales, as in Eqs.~(\ref{eq:C-sep}) and (\ref{eq:R-sep}).
This {\it Ansatz} is introduced in the dynamic equations for $C$ and $R$, Eqs.~(\ref{eq:dyn-eqs-C})-(\ref{eq:dyn-eqs-z}), and the
evolution in two two-time sectors is studied separately together with the requirement
that the behaviour matches in the crossover region. The resulting equations are manipulated a bit,
and equations for the parameters $q$, $T_f$ and $T_{\rm eff}$, are derived. We reckon that with
this procedure we introduce three unknowns and we deduce five equations, one being
the energy conservation. The other four equations are
the equations for $C_{\rm st}$, $C_{\rm ag}$, $R_{\rm st}$ and $R_{\rm ag}$,
but these are not all independent, since the FDT with $T_f$ for ($C_{\rm st}$, $R_{\rm st}$) and $T_{\rm eff}$ for
$(C_{\rm ag}$, $R_{\rm ag}$) reduce their number to two.  There are then three unknowns and three
equations.

\subsection{Stationary dynamics}
\label{subsec:stationary}

In this Section we derive the set of equations that determine $q,q_0$ and $T_f$ as a function of the properties of the initial state, $T'$,
$m_0$ and $J_0$,
and the ones after the quench, $m$ and $J$, assuming that a stationary state is reached.

The stationarity assumption implies
\begin{equation}
\lim_{t_1\to\infty} z(t_1) = z_f
\; ,
\qquad
 \lim_{t_2\to\infty} C(t_1, t_2) = C_{\rm st}(t_1-t_2)
\; ,
\qquad
 \lim_{t_2\to\infty} R(t_1, t_2) = R_{\rm st}(t_1-t_2)
\; .
\end{equation}
where we took $t_1 \geq t_2$.
The large $\tau \equiv t_1 - t_2$ limit can then be further considered to define
\begin{equation}
q\equiv \lim_{\tau \to \infty} C_{\rm st}(\tau)
\; .
\end{equation}
This asymptotic limit has to be distinguished from the one of the correlation between the initial condition and
the dynamic configuration
\begin{equation}
q_0 \equiv \lim_{t_1\to\infty} C(t_1,0^+)
\; .
\end{equation}
The parameters $q$ and $q_0$ will take zero or non-vanishing values in different situations presented below.
Accordingly, $C_{\rm st}(0) \neq C(0,0)$ since in the former equation the $t_2\to\infty$ limit has been taken and
in the latter equation $t_2=0^+$. In the following presentation we drop the superscript $+$
from the initial time but the $0$ of the  absolute times should be understood as~$0^+$.

If $R_{\rm st}$ and $C_{\rm st}$ satisfy FDT with respect to a temperature $T_f$, and we call $\tau=t_1-t_2$,
\begin{equation}
R_{\rm st}(\tau) = - \frac{1}{T_f} \; d_\tau C_{\rm st}(\tau)
\qquad\qquad
\mbox{and}
\qquad\qquad
\chi_{\rm st}(\tau) \equiv \int_0^\tau d\tau' \ R_{\rm st}(\tau') = \frac{1}{T_f} [1-C_{\rm st}(\tau)]
\; ,
\label{eq:chi-st}
\end{equation}
where we used $C_{\rm st}(0)=1$.
The second way of writing the FDT is the one that we will
exploit in the numerical analysis to determine the final temperature $T_f$
from the plot of $\chi_{\rm st}$ against $C_{\rm st}$ constructed using the time-lag
$\tau$ as a parameter.

In order to make the presentation of the analytic part easier we list here
the steps followed in the derivation of the asymptotic equations under the stationary assumption:\\
\indent
- We take the asymptotic limit of the equation for $z(t_1)$ and write $z_f \equiv \lim_{t_1\to \infty} z(t_1)$ as a function of the

$\;\; q_0$ and $q$
parameters and $T_f$.
\\
\indent
- We write the conservation of the energy.
\\
\indent
- We prove that the equation for $R_{\rm st}$ becomes the $\tau$-derivative of the $C_{\rm st}$ equation.
\\
\indent
- We take the asymptotic limit of the equations for $C(t_1,0)$ and $C_{\rm st}(\tau)$.
\\
The conservation of the total energy and the two last equations derived
constitute a set that fixes $T_f$, $q_0$ and $q$ knowing $T'$, $J_0$, $J$, $m_0$ and $m$.
We do not prove analytically that the asymptotic solution is reached by the dynamics, this would need the full solution of the
equations of motion and a matching problem that remains out of reach analytically.  In contrast,
we do verify {\it a posteriori} for which set of parameters $(T', J_0, J, m_0, m)$ this occurs by solving numerically the full set of equations.

The steps followed in the case in which the system ages and stationarity is broken are rather similar but need some
generalisation, see Sec.~\ref{subsec:nonstationary}.

\subsubsection{The asymptotic Lagrange multiplier and the total energy}
\label{subsec:energies}

Starting from Eq.~(\ref{eq:z-Nicolas}) and using FDT
\begin{equation}
z(t_{1})=2e_f+  \frac{ J^2 p \left(p+2\right)}{2} \int_{0}^{t_{1}}dt' \; \frac{1}{T_f} \left( \partial_{t'} C(t_{1},t') \right)
C^{p-1}(t_{1},t')+ \frac{JJ_0 (p+2)}{2T'} \ C^{p}(t_{1},0)
\end{equation}
the integral can be computed and an asymptotic expression for $z_f$ is obtained
\begin{equation}
\displaystyle{z_f =
2 e_f
+\frac{J^2 (p+2)}{2T_f} (1-q^p) + \frac{JJ_0(p+2)}{2T'} q_0^p
}
\; .
\label{eq:eq-zinf-extraction}
\end{equation}

Proceeding similarly, the potential energy is given by
\begin{equation}
e^f_{\rm pot} = - \frac{J^2}{2T_f} \, (1-q^p) - \frac{JJ_0 }{2T'} \, q_0^p
\; ,
\label{eq:epot-finiteq}
\end{equation}
that becomes the paramagnetic result $e^f_{\rm pot} = - J^2/(2T_f)$  for $q_0=q=0$.
Moreover, if $J=J_0$ and $T_s \leq T'=T_f$ (no potential energy quench), $q_0=q$ and  $e^f_{\rm pot} = -J^2/(2T_f)$, independently of
$q$, as it should. Note that we need the contribution from the last term to get the correct no-quench limit.
Besides, we assume that the asymptotic kinetic energy is determined by ``equipartition'' at the final temperature
\begin{equation}
e^f_{\rm kin} = \frac{T_f}{2}
\; .
\end{equation}
Then, the asymptotic total energy reads
\begin{equation}
\displaystyle{e_f=e^f_{\rm kin} + e^f_{\rm pot} = \frac{T_f}{2} - \frac{J^2}{2T_f} \, (1-q^p) - \frac{JJ_0 }{2T'} \, q_0^p} \; .
\label{eq:total-energy}
\end{equation}


We argued that the energy right after a quench in the interactions and mass is $e_f=e(0^+) = mT'/(2m_0) - J J_0/(2T')$. Compared to the
asymptotic form derived in Eq.~(\ref{eq:total-energy}) the conservation of the total energy implies
\begin{equation}
\frac{mJ_0}{m_0J} \frac{T'}{T_0} - \frac{J_0}{T'} = \frac{T_f}{J} - \frac{J}{T_f} \, (1-q^p) - \frac{J_0 }{T'} \, q_0^p
\; .
\label{eq:energy-simplified}
\end{equation}
We see here two adimensional parameters $T'/J_0$ and $Jm_0/(J_0m)$ that characterise the pre-quench conditions and the
comparison between the pre and post quench parameters.

The equation for $z_f$ in Eq.~(\ref{eq:eq-zinf-extraction}) can now be rewritten as
\begin{equation}
z_f = T_f + \frac{J^2p}{2T_f} (1-q^p) + \frac{JJ_0p}{2T'} q_0^p
\end{equation}
after replacing the energy $e_f$ by its dependence on $T_f$. It takes now a form that is very similar to the one of
the relaxation dynamics~\cite{CuKu93}.

\subsubsection{The asymptotic analysis of the correlation equation}
\label{subsec:derivation}

The equation for $C$ can be treated in two regimes of times.
In one case we take $\tau=t_1-t_2$ fixed and $t_1$ and $t_2$ tending to infinity. The equation for $C$ then reads
\begin{eqnarray}
[md_{\tau^2} + z_f] \, C_{\rm st}(\tau)
&=&
\frac{J^2 p }{2T_f} \; \int_0^{t_2\to\infty} dt' \frac{\partial}{\partial t'} \left( C_{\rm st}^{p-1}(t_1,t') C_{\rm st}(t_2,t')\right)
\nonumber\\
&& + \frac{J^2p}{2T_f} \; \int_{0}^{\tau} d\tau' \; \left(\partial_{\tau'} C_{\rm st}^{p-1}(\tau-\tau') \right) C_{\rm st}(\tau')
+ \frac{JJ_0p}{2T'} C^{p-1}(t_1,0) C(t_2,0)
\; .
\end{eqnarray}
When deriving this equation we assumed that the contribution to the integrals
of any possible transient between the time $0$ and a time $t_{\rm tr}$ after which
the FDT establishes can be neglected. The lower limit $0$ in the integral over $t'$ is then to be
interpreted as the initial time $t_{\rm tr}$ of this asymptotic regime, although we simply write
$0$ in these equations. In the second integral $0$ is the minimal time-delay at which the functions $C_{\rm st}$
and $R_{\rm st}$ are measured.

We can treat in the same way the equation for $R$ and then compare the two. As already mentioned
in the list that summarizes the steps to follow, we can prove that the equation for $R_{\rm st}(\tau)$ is the time-delay derivative of the equation for
$C_{\rm st}(\tau)$ times $1/T_f$. This is a quite straightforward calculation that we choose not to show here.

In the limit $t_1 \geq t_2 \to \infty$ we can replace $C(t_1,0)$ and $C(t_2,0)$ by $q_0$.
We further take the limit $\tau\to \infty$ and drop the second time derivative assuming that the dynamics become
slow at long time delays. The first integral
is computed as it is written now. The second one is made more symmetric before
approximating it, in such a way that the two extremes ($0$ and $\tau$) contribute in the same
way. One has
\begin{eqnarray}
 z_f q
&=&
\frac{J^2 p }{2T_f} \;  (q^{p-1}-q^{p})
\nonumber\\
&&
+ \frac{J^2p}{2T_f} \;
\lim_{\tau\to\infty} \int_{0}^{\tau} d\tau' \, \left[ \frac{1}{2} (d_{\tau'} C^{p-1}(\tau-\tau')) C(\tau')
+ \frac{1}{2} d_{\tau'} (C^{p-1}(\tau-\tau')  C(\tau') ) - \frac{1}{2} C^{p-1}(\tau-\tau')   d_{\tau'} C(\tau')
\right]
\nonumber\\
&&
+ \frac{JJ_0p}{2T'} \; q_0^p
\nonumber\\
&=&
\frac{J^2 p }{2T_f} \; (q^{p-1} -q^{p})
+ \frac{J^2p}{2T_f} \;
\frac{1}{2} \left[ q (1-q^{p-1}) + \frac{1}{2} (q-q^{p-1}) - \frac{1}{2} q^{p-1} (q-1) \right]
 + \frac{JJ_0p}{2T'} \; q_0^p
 \; .
 \end{eqnarray}
 Finally,
\begin{equation}
\displaystyle{z_f q=
\frac{J^2 p }{2T_f} \; (q^{p-1} -q^{p})
+ \frac{J^2p}{2T_f} \;
q (1-q^{p-1})
+ \frac{JJ_0p}{2T'} \; q_0^p
}
\; .
\label{eq:eq-q-extraction}
\end{equation}
This equation admits the solution $q_0=q=0$ but it can also have, for certain values of the parameters, solutions with $q\neq 0$
and $q_0\neq 0$ being equal or different.

The other interesting limit is the one in which we set $t_2$ to be strictly $0$ and we tend $t_1$  to infinity. The equation for $C$ becomes
\begin{eqnarray}
[md_{t_1^2} + z_f] \, C(t_1,0)
=
\frac{J^2p(p-1)}{2} \; \int_{0}^{t_1} dt' \; R(t_1,t') C^{p-2}(t_1,t') C(t',0)
+ \frac{JJ_0p}{2T'} C^{p-1}(t_1,0) C(0,0)
\; .
\end{eqnarray}
In the limit $t_1\to\infty$ we can replace $C(t_1,0)$ by $q_0$ and use $C(0,0)=1$.
We further drop the second time derivative, and use stationarity and FDT, to find
\begin{eqnarray}
 z_f q_0
&=&
 \displaystyle{
 \frac{J^2 p }{2T_f} \;  q_0 (1-q^{p-1})
+ \frac{JJ_0p}{2T'} \; q_0^{p-1}
}
\; .
\label{eq:eq-q0-extraction}
\end{eqnarray}
This equation admits the solution $q_0=0$ but it can also have, for certain values of the parameters, solutions with $q_0\neq 0$.
As a check of consistency, we remark that for $J=J_0$ and $T'=T_f$ the two remaining equations, Eqs.~(\ref{eq:eq-q-extraction}) and~(\ref{eq:eq-q0-extraction}), are
compatible for $q=q_0$.

We can now write down two other equations that relate $q_0, q$ and $T_f$:
\begin{eqnarray}
\left[
T_f
+\frac{J^2 p}{2T_f}  (1-q^p) + \frac{JJ_0 p}{2T'} q_0^p
\right]
q_0
 &=& \frac{J^2 p }{2T_f} \;  (1-q^{p-1}) \, q_0
+ \frac{JJ_0p}{2T'} \; q_0^{p-1}
\; ,
\label{eq:qq-simplified}
\\
\left[
T_f
+\frac{J^2 p}{2T_f}  (1-q^p) + \frac{JJ_0p}{2T'} q_0^p
\right] q &=&
\frac{J^2 p }{2T_f} \; q^{p-1} (1- q)
+ \frac{J^2p}{2T_f} \;
q (1-q^{p-1})
+ \frac{JJ_0p}{2T'} \; q_0^p
\; .
\label{eq:qq0-simplified}
\end{eqnarray}

\subsubsection{The equations fixing the parameters $q$, $q_0$ and $T_f$}

After some rearrangements, the three equations (\ref{eq:energy-simplified}), (\ref{eq:qq-simplified}) and
(\ref{eq:qq0-simplified}) simplify to
\begin{eqnarray}
\frac{J_0m}{Jm_0} \frac{T'}{J_0} - \frac{T_f}{J}
&=&
- \frac{J}{T_f} \, (1-q^p) + \frac{J_0 }{T'} \, (1-q_0^p)
\; ,
\label{eq:energy-simplified2}
\\
q_0 \; \frac{T_f}{J}
  &=& - \frac{p}{2} \frac{J}{T_f}  q_0 q^{p-1} (1-q)  + \frac{p}{2} \frac{J_0}{T'} \; q_0^{p-1} (1-q_0^2)
  \; ,
  \label{eq:q0-simplified}
\\
q \; \frac{T_f}{J}
&=&
\frac{p}{2}
\frac{J}{T_f} \; q^{p-1}
(1-q)^2
+ \frac{p}{2} \frac{J_0}{T'} \; q_0^p \; (1-q)
\; .
\label{eq:q-simplified}
\end{eqnarray}
One can use Eqs.~(\ref{eq:energy-simplified2}), (\ref{eq:q0-simplified}) and (\ref{eq:q-simplified}) to determine $q_0, q, T_f$ in situations in
which a steady state is reached.

More simplifications are possible if one extracts $q^p (1-q)/q$ from the second equation and inserts it in the third one
to obtain
\begin{equation}
\frac{T_f}{J} \frac{T'}{J_0}= \frac{p}{2} \ q_0^{p-2} (1-q)
\label{eq:dos}
\end{equation}
a linear equation in $q$.

We can now check that for $J=J_0$ and $m=m_0$, the equation that expresses energy conservation is consistent with $T'=T_f$ and $q=q_0$.
Moreover, taking $q_0=q$, Eqs.~(\ref{eq:q-simplified}) and (\ref{eq:dos}) become the same one,
\begin{equation}
\frac{p}{2} \ q^{p-2} (1-q) = \left(\frac{T_f}{J}\right)^2
\; ,
\label{eq:solq0=q}
\end{equation}
that is the equation for $q$ in the TAP solutions that are mixed to yield the non-trivial paramagnet  at $T^0_s < T' <T^0_d$, see Eq.~(\ref{eq:q-eq-no-quench}). The dynamics remain confined in the initial TAP state where the system was prepared.

\subsubsection{Quench dynamics, target paramagnetic state}

Let us look for solutions with $q_0=q=0$ that correspond to a final paramagnetic
state. Equations (\ref{eq:q0-simplified}) and (\ref{eq:q-simplified}) are identical to zero and
Eq.~(\ref{eq:energy-simplified2}) implies
\begin{equation}
\frac{J_0m}{Jm_0} \frac{T'}{J_0}- \frac{J_0}{T'}  = \frac{T_f}{J}  - \frac{J}{T_f}
\end{equation}
that fixes $T_f$
\begin{equation}
T_f^{(1,2)} = J \; \frac{J_0}{2T'}
\left[ \frac{J_0m}{Jm_0} \left(\frac{T'}{J_0}\right)^2 - 1  \pm \sqrt{\left[\frac{J_0m}{Jm_0} \left(\frac{T'}{J_0} \right)^2 - 1\right]^2 +  \left(\frac{2T'}{J_0}\right)^2} \right]
\; .
\label{eq:eq_temp_x}
\end{equation}
As the temperature cannot be negative, the plus sign is the relevant one here.
This relation can be used to check whether the system has really attained thermal equilibrium by looking at the
parametric plot of the integrated linear response, $\chi(t_1,t_2) = \int_{t_1}^{t_2} dt' \ R(t_1, t')$, as
a function of the correlation function, $C(t_1,t_2)$, and comparing minus the inverse slope with $T_f$.
For a stationary system, the expected linear form is given in Eq.~(\ref{eq:chi-st}).

The  temperature in the asymptotic paramagnetic state, in units of $J$,
is a function of $T'/J_0$ and $J m_0/(J_0 m)$.
One can easily show from the analytic form above that
$T_f= J (J_0m/(Jm_0) \, (T'/J_0)$ at $m_0J/(mJ_0)\to 0$. In the limit $Jm_0/(J_0m)\to\infty$ the temperature approaches
$T_f\to J/(2T') (-J_0 + \sqrt{J_0^2 + 4 {T'}^2})$. If, moreover, $T'/J_0$ also diverges, it becomes
$T_f \to J$. Some special finite values are
$T_f=JT'/J_0$ for $m_0 J/(mJ_0)=1$, and $T_f=T'$ if $m_0=m$
and $J_0=J$.

The curve $T_f/J$ as a function of the full control parameter $(Jm_0/(J_0m))$ is shown with dotted blue lines
in the two panels in Fig.~\ref{fig:temps} that represent quenches from  equilibrium at $T'=0.7 J_0$ and $T'=0.6 J_0$.
The open circles correspond to dynamical runs that realise the paramagnetic asymptotic solution. We will discuss the region of
parameters in which this is the asymptotic state in Secs.~\ref{sec:numerical}~and~\ref{sec:phase-diagram}.

The relation between the final total energy and the final temperature is very simple in the paramagnetic state
\begin{equation}
T_f = e_f + \sqrt{e_f^2 + J^2 }
\; ,
\label{eq:eq_temp_pm}
\end{equation}
and it coincides with the result of inverting Eq.~(\ref{eq:dyn_en}), that is to say, the equilibrium paramagnetic energy
as a function of  temperature.

\subsubsection{Dynamics within metastable states}
\label{subsec:dyn_metastable}

When the temperature of the initial condition is such that $T^0_s<T'<T^0_d$, the system is prepared in a TAP state. In this Section we will show that
whenever the asymptotic equations~(\ref{eq:energy-simplified2})--(\ref{eq:q-simplified}) have solutions with $q,q_0\neq 0$, the dynamics of the system
are confined to the same TAP state that, after the change in the coupling strength operated at the quench,
is only translated in the potential energy landscape and possibly rescaled in size, thanks to the fact that in the
spherical $p$-spin model there is no birth, death or merging of states at intermediate temperatures.

The first remark is that, as already mentioned, the interaction quench changes the depth of the potential energy minima.
If one minimum has energy $e_{\rm pot}^{0}$ initially, its energy after the quench is given by
\begin{equation}
e_{\rm pot}=\frac{J}{J_0}e_{\rm pot}^{0}
\; .
\label{eq:follow-e-TAP}
\end{equation}

Given that $T^0_s<T'<T^0_d$, the initial state is described by Eq.~(\ref{eq:q-eq-no-quench}) that we here rewrite
making the dependence of the parameter $q$ on the initial temperature, $T'$,
and strength of the random potential, $J_0$,
\begin{equation}
\frac{J^2_0 p}{2T'^2} \ q[J_0,T']^{p-2}(1-q[J_0,T'])=1
\; ,
\label{eq:initial_q}
\end{equation}
explicit.
Indeed,  $q[J_0,T']$ is the equilibrium value of the self overlap at the initial temperature $T'$.
This equation  can be written in a slightly different way that will be useful later
\begin{equation}
q[J_0,T']-\left[\frac{J^2_0 p}{2}(1-q[J_0,T'])\right]^{-\frac{1}{p-2}}(T')^{\frac{2}{p-2}}=0
\; .
\label{eq:initial_q2}
\end{equation}

We call $e^{0}_{T'}$ the bare potential energy of the TAP state that dominates the partition function at the initial temperature $T'$.
In such case, from Eq.~(\ref{eq:tap}), $q_{T'}$ also fullfills
\begin{equation}
(p-1)(1-q[J_0,T'])q[J_0,T']^{(p-2)/2}=\frac{T'}{J^2_0}\left[-e_{T'}^{0}-\sqrt{(e^0_{T'})^2-e^{0}_c}\right]
\; ,
\label{eq:initial_tap}
\end{equation}
where $e^{0}_{c}=-J_0\sqrt{2(p-1)/p}$. According to Eq.~(\ref{eq:follow-e-TAP}), after the quench, the energy of this
TAP state is given by $e_{T'}= (J/J_0) \ e_{T'}^{0}$.

Let us call $q[J,T_f]$ the self overlap in the final TAP state at temperature $T_f$. Also from Eq.~(\ref{eq:tap}), and since the energy of the TAP state is rescaled, it is clear that $q[J,T_f]$ satisfies
\begin{equation}
(p-1)(1-q[J,T_f])q[J,T_f]^{(p-2)/2}=\frac{T_f}{J^2}\left[-e_{T'}-\sqrt{e^2_{T'}-e_c}\right]
\; ,
\end{equation}
where $e_{c}=-J\sqrt{2(p-1)/p}$. Recalling that $e_{T'}=(J/J_0) \ e_{T'}^{0}$, we can write
\begin{equation}
(p-1)(1-q[J,T_f])q[J,T_f]^{(p-2)/2}=\frac{T_f}{JJ_0}\left[-e_{T'}^{0}-\sqrt{(e^0_{T'})^2-e^{0}_c}\right]
\; .
\label{eq:final_tap}
\end{equation}
Then, from Eqs.~(\ref{eq:initial_q}),~(\ref{eq:initial_tap}) and~(\ref{eq:final_tap}) we obtain a relation between $q[J_0,T']$ and $q[J,T_f]$
\begin{equation}
q[J_0,T']=1-\frac{J^2 p}{2T^2_f}(1-q[J,T_f])^2q[J,T_f]^{p-2}
\; .
\label{eq:relation_if}
\end{equation}
Using now the results for the asymptotic analysis of the dynamical equation, more precisely,
Eqs.~(\ref{eq:q-simplified}) and~(\ref{eq:dos}),
 the equation for $q=\lim_{\tau \to \infty} C(\tau)$ can be written as
\begin{equation}
1=\frac{J^2p}{2T^2_f}q^{p-2}(1-q)^2+\left[\frac{JJ_0p}{2T'T_f}\right]^{-\frac{2}{p-2}}q^{-1}(1-q)^{-\frac{2}{p-2}}
\; .
\label{eq:intermediate}
\end{equation}
 Defining
\begin{equation}
U \equiv 1-\frac{J^2p}{2T^2_f}q^{p-2}(1-q)^2
\; ,
\end{equation}
we can write Eq.~(\ref{eq:intermediate}) as
\begin{equation}
U-\left[\frac{J^2_0 p}{2}(1-U)\right]^{-\frac{1}{p-2}}(T')^{\frac{2}{p-2}}=0
\; .
\label{eq:key}
\end{equation}
Comparing this last equation with Eq.~(\ref{eq:initial_q2}) we conclude that
\begin{equation}
U=q[J_0,T']
\; ,
\end{equation}
which, using Eq.~(\ref{eq:relation_if}), implies
\begin{equation}
q=q[J,T_f]
\; .
\end{equation}
This shows that the system remains trapped in the same metastable state during all the evolution. Of course,
if the quench takes the system to parameters (final temperature $T_f$) such that this state does no
longer exist, the system escapes it into the proper paramagnetic state.

In Fig.~\ref{fig:temps}~(b) we draw, with blue dotted lines, the $J m_0/(J_0 m)$ dependence of $T_f$
for the asymptotic TAP states and the open squares show the numerical solution of the full equations
for two choices of $Jm_0/(J_0 m)$ that realise this asymptotic state.

\subsection{Non stationary dynamics and ageing}
\label{subsec:nonstationary}


Let us now explain how the ageing equations are studied.
In the aging regime we expect the correlation with the initial configuration to decay to zero
\begin{equation}
\lim_{t_1\to\infty} C(t_1,0) = 0
\; .
\end{equation}
The dynamic equations (\ref{eq:dyn-eqs-C})-(\ref{eq:dyn-eqs-z})
therefore lose the terms that depend on the initial conditions. The only formal difference with the
equations for the dissipative case~\cite{CuKu93,Ba97} is
that the friction term (first time derivative) is now replaced by the inertial term (second-time derivative)
and that the temperature is not fixed {\it a priori}.

\subsubsection{The parameters $q$, $T_f$, and $T_{\rm eff}$}

Following the explanation in~\cite{CuKu93}, explained in more detail in~\cite{LesHouches}, the study of the $C$ and $R$
yields the equation that fixes plateau parameter $q$ to be the one
on the threshold, Eq.~(\ref{eq:qth}). In the stationary regime the temperature is given by the parameter $T_f$ in the FDT linking
$C_{\rm st}$ and $R_{\rm st}$, that is not fixed yet. Moreover, the combination of the $C$ and $R$ equations in the stationary and
aging regime yields the equation that fixes the effective temperature in the aging regime, $T_{\rm eff}$,  and this equation is,
again, the same as in the dissipative case, Eq.~(\ref{eq:mth}). We have
\begin{eqnarray}
 \frac{T_f^2}{J^2} &=& \frac{p(p-1)}{2} \ q^{p-2} (1-q)^2
\; ,
\label{eq:q-const-energy}
\\
\frac{T_f}{T_{\rm eff}} &=& \frac{(p-2)(1-q)}{q}
\; .
\label{eq:Teff-const-energy}
\end{eqnarray}
It is not
necessary to fix the value of the Lagrange multiplier to derive these two equations.
We still need to find, though, which is the value of $T_f$ selected by the closed system.

The selection of $T_f$ is done by the energy conservation. The asymptotic energy is the sum of the kinetic contribution,
$T_f/2$, and the potential one that reads
\begin{equation}
e_{\rm pot}^f = - \frac{J^2}{2} \left[ \frac{1}{T_f} (1-q^p) + \frac{1}{T_{\rm eff}} q^p
\right]
\; .
\label{eq:pot_aging}
\end{equation}
Therefore
\begin{equation}
\frac{J_0m}{Jm_0} \frac{T'}{J_0}- \frac{J_0}{T'} =  \frac{T_f}{J} - \left[ \frac{J}{T_f} (1-q^p) + \frac{J}{T_{\rm eff}} q^p \right]
\; .
\label{eq:ef-const-energy}
\end{equation}
We now have three equations for the three unknowns $q, \ T_f$ and  $T_{\rm eff}$. These equations can
be simplified and recast in a more convenient manner by replacing the $q$-dependence of
$T_f/T_{\rm eff}$ from Eq.~(\ref{eq:Teff-const-energy}) in Eq.~(\ref{eq:ef-const-energy}), that is now a
quadratic equation on $T_f/J$. Solving for $T_f/J$ and replacing the result in Eq.~(\ref{eq:q-const-energy}),
after a straightforward calculation,
we obtain
\begin{equation}
\left(\frac{J_0m}{Jm_0} \frac{T'}{J_0}- \frac{J_0}{T'} \right)^2 =
 \frac{2 \ q^{2-p}}{p(p-1) (1-q)^{2}}
 \left[
 \frac{p(p-1)}{2} q^{p-2} (1-q)^2 - (1-q^p) - (p-2)(1-q) q^{p-1}
 \right]^2
 \; .
 \label{eq:q-good}
\end{equation}
Equation~(\ref{eq:q-good}) determines $q$ given the initial temperature $T'$,
the pre and post quench variance of the random interactions parametrized by $J_0$ and $J$
and the pre and post quench masses $m_0$ and $m$. These parameters appear in the combinations
$T'/J_0$ and $J m_0/(J_0 m)$.
Once $q$ is found, Eqs.~(\ref{eq:q-const-energy}) and (\ref{eq:Teff-const-energy}) yield $T_f/J$ and $T_{\rm eff}/J$, respectively.

The solutions to Eq.~(\ref{eq:q-good}) can be understood graphically. The r.h.s. is a function of $q$ with positive curvature and
a single minimum at $q=(p-2)/p$, the overlap at the spinodal,
in the interval $q\in[0,1]$. The equation has two solutions, the one with larger value being the relevant one.
When the control parameter $m_0 J/(m J_0)$ is decreased, the
equation ceases to have solution at
\begin{equation}
\left(\frac{m_0J}{mJ_0}\right)_{\rm min} = \frac{T'}{J_0} \frac{1}{\displaystyle{\frac{J_0}{T'}} + \sqrt{\mbox{r.h.s. Eq.~(\ref{eq:q-good})} \left(\frac{p-2}{p} \right)}}
\; ,
\end{equation}
when the l.h.s. touches the minimum of the r.h.s. This value is $\approx 1.295$ for $T'/J_0=0.7$ and $\approx 0.77$ for $T'/J_0=0.6$,
and $p=3$ (see the ending points of the ageing $T_f/J$ and $T_{\rm eff}/J$ curves in Fig.~\ref{fig:temps}, and the discussion
of the phase diagram in Sec.~\ref{sec:phase-diagram}).


In the ageing solutions, as long as $T_f$ (the  temperature of the fast degrees of freedom) is lower
than $T_d$, the effective temperature is larger than $T_f$. When $T_f$ goes beyond $T_d$, its relation with the
effective temperature is inverted, and it becomes larger than $T_{\rm eff}$:
\begin{eqnarray}
&& T_{\rm eff}>T_f \qquad \mbox{for} \qquad T_d>T_f
\nonumber\\
&& T_{\rm eff}<T_f \qquad \mbox{for} \qquad T_d<T_f
\; .
\end{eqnarray}
The curves of $T_f$ and $T_{\rm eff}$, as functions of $Jm_0/(J_0m)$ are shown in Fig.~\ref{fig:temps} with
solid grey and dotted grey lines, respectively. The open triangles indicate the actual solution
of the full equations found numerically. The temperature inversion, $T_f>T_{\rm eff}$ predicted by the asymptotic
ageing equations for $T_f>T_d$ is not realised asymptotically by the full equations but, as we will show in
Sec.~\ref{subsubsec:TAP-to-spinodal}, it appears in a transient regime.

\subsection{Summary of asymptotic solutions}

\begin{figure}[h]
\centerline{
(a) $T'=0.7 \, J_0>T_d^0 \qquad\qquad\qquad\qquad\qquad\qquad\qquad\quad$ (b) $T'=0.6 \, J_0 <T_d^0$
}
\vspace{0.25cm}
\centerline{
\includegraphics[scale=0.7]{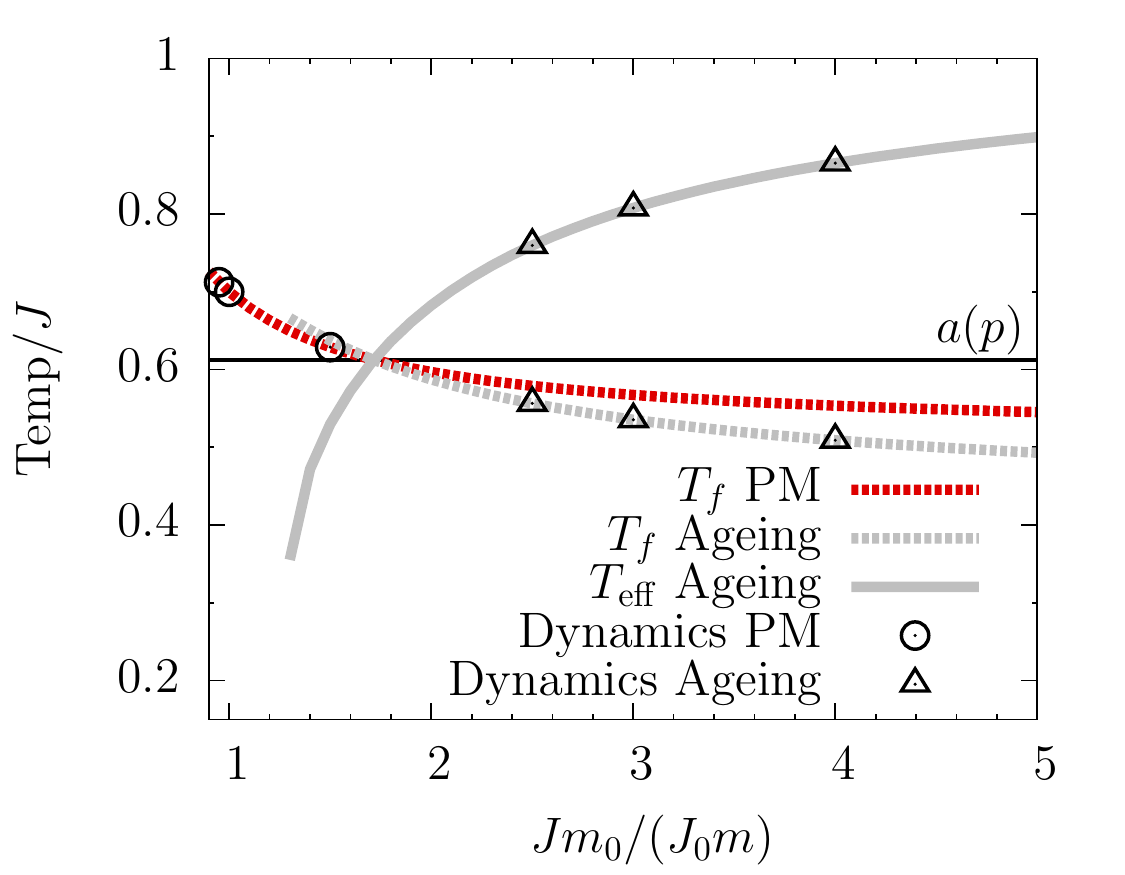}
\includegraphics[scale=0.7]{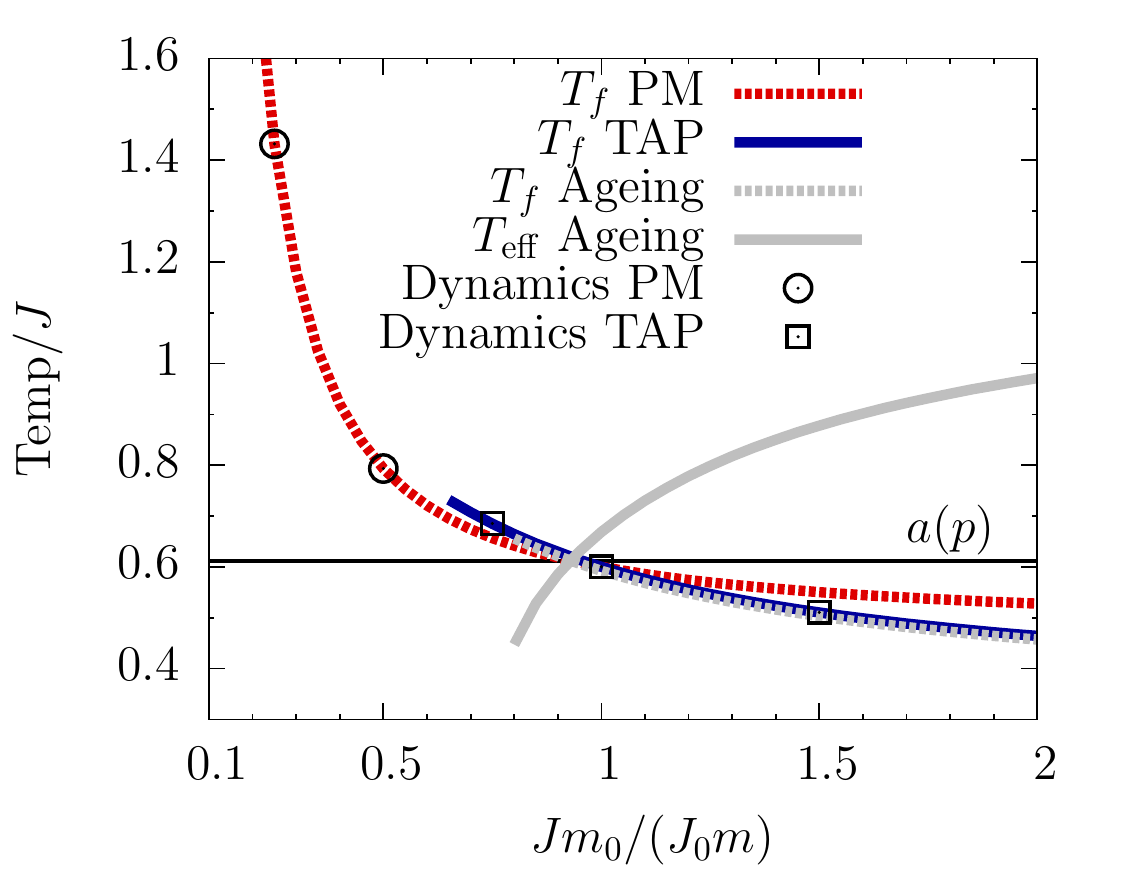}
}
\caption{\small Characteristic temperatures in units of $J$ after quenches from
(a) an initial  paramagnetic state with $T'=0.7 \, J_0 >T_d^0$
and (b) an initial TAP state with $T'=0.6 \, J_0<T_d^0$. The lines represent the
asymptotic solutions as indicated in the keys. The data points are the
results of the numerical solution of the full set of equations. For each set of pre and post quench
parameters one and only asymptotic state is realised.
}
\label{fig:temps}
\end{figure}

As a summary of the different asymptotic solutions, we show in Fig.~\ref{fig:temps}  the
$m_0J/(mJ_0)$-dependence of the dynamic critical temperature, $T_d$ from Eq.~(\ref{eq:Td}), the final temperature $T_f$ of
the PM ($q=q_0=0$) and TAP ($q,q_0\neq0$) branches of
the solutions to Eqs.~(\ref{eq:energy-simplified2})--(\ref{eq:q-simplified}) (stationary \textit{Ansatze}),
and the final temperature, $T_f$, and effective temperature, $T_{\rm eff}$, derived from the solution to the
set of equations (\ref{eq:q-const-energy})-(\ref{eq:q-good}) (ageing \textit{Ansatze}).
We note that the ageing temperatures $T_f$ and $T_{\rm eff}$ coincide with $T_d$ at a single value of the control parameter
$Jm_0/(J_0m)$.
We also show with different points the results from the full solution of the evolution
equations, indicating which is the asymptotic solution realised by the dynamics in
each range of parameters.

In panel (a),  the initial state is paramagnetic $T'=0.7 J_0>T_d^0$,
and the dynamics choose the PM solution (red dotted line and open triangles) for energy injection
or for small energy extraction, while for sufficiently large energy extraction the asymptotic dynamics show ageing,
characterised by two temperatures (grey lines).

In panel (b), $T'=0.6 J_0<T_d^0$ and the initial configurations are drawn within a TAP state.
The dynamics choose the PM solution (red dotted) for large energy injection, while the system
remains in a TAP state (blue lines and open triangles) for small energy injection or for energy extraction.
The ageing solution (grey lines) are not realised by the dynamics.

The exact boundaries between the different kinds of solutions
selected by the full equations will be derived analytically in Sec.~\ref{sec:phase-diagram}.
In general, transients appear  in the parameter regions where the system changes its asymptotic behavior.

\vspace{0.5cm}

\section{Numerical results}
\label{sec:numerical}

In the numerical solution of the full set of equations we fix $J_0=m_0=m=1$. This means that the initial energy landscape is fixed. In particular,
we know the values of the initial critical temperatures $T^0_d=0.612$ and $T^0_s=0.586$.
We shall then vary the initial temperature $T'$ and the coupling $J$ of the Hamiltonian that drives the time evolution.

For later reference and recalling the discussion in Sec.~\ref{sec:quenches},
the critical temperatures corresponding to the equilibrium landscape with the final coupling $J$ can be calculated
by noticing that the critical temperatures are proportional to the coupling. Then
\begin{equation}
\frac{T_s}{T_s^0} = \frac{J}{J_0} \qquad \Rightarrow \qquad
T_s = \frac{J}{J_0} \ T^0_s
\qquad\mbox{and}\qquad
T_d = \frac{J}{J_0} \ T^0_d
\; .
\end{equation}

After some general considerations about the numerical algorithm we analyse some specific processes to illustrate  the
analytical results of the previous Section and put them to the test.
We will consider energy injection and energy extraction  processes sketched in Figs.~\ref{fig:up} and \ref{fig:down}, respectively.
The full numerical solution to the equations allows to prove which among the asymptotic solutions are realised, when several
co-exist.

\subsection{Equilibrium dynamics}

We first checked that for $J=J_0$ and $m=m_0$, that is to say $\Delta e=0$, the system has a stationary evolution for all equilibrium
initial conditions.

We studied the no energy change case with two purposes. One is to check consistency of our numerical algorithm. The other is to
investigate the effect of the discretisation step $\delta$ on the results obtained from the numerical integration of the
equations. We found that the algorithm does conserve energy and that a step $\delta=0.0025$ was sufficient to assure numerical convergence
of our results.

\vspace{0.25cm}

\begin{figure}[h]
\begin{center}
\includegraphics[scale= 0.6]{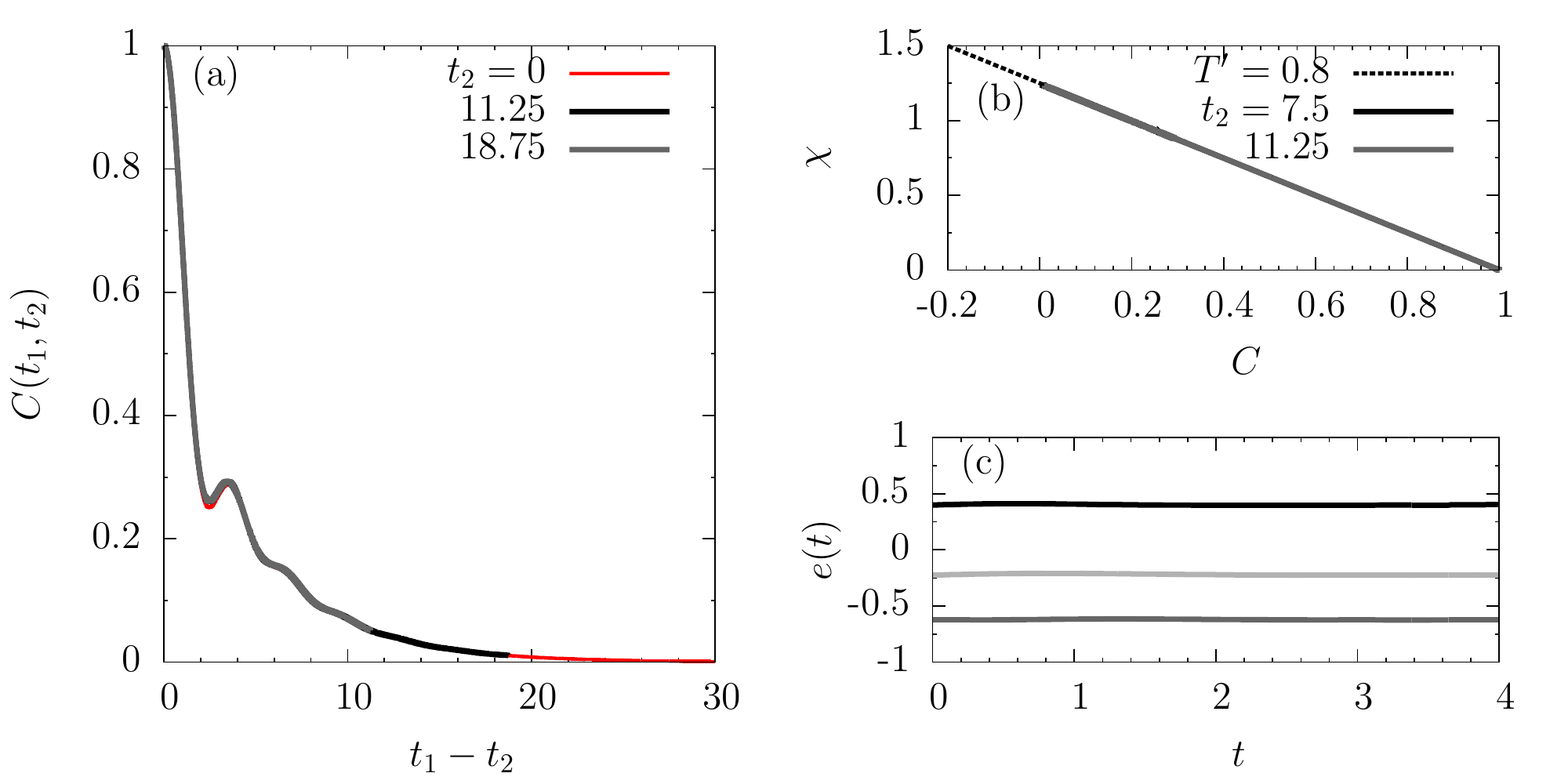}
\end{center}
\caption{\small Constant energy dynamics, $\Delta e=0$ ($J=J_0$ and $m=m_0$). $T'=0.8>T^0_d$ in the paramagnetic phase.
(a) Dynamics of the correlation function for various choices of the waiting time given in the key. The stationary relaxation to zero
is clear.  (b) Linear-response vs. correlation parametric plot for two values of the waiting time $t_2$ indicated in the plot.
The dashed line shows the FDT with the initial temperature. (c) Energy time-dependence.
The data correspond to $e_{\rm kin}$ (above), $e_f$ (middle), and $e_{\rm pot}$ (below).
The numerically evaluated values of the potential and kinetic energies and temperature
agree with the ones derived analytically within numerical accuracy.
}
\label{fig:T08_eq}
\end{figure}

\begin{figure}[h]
\begin{center}
\includegraphics[scale= 0.6]{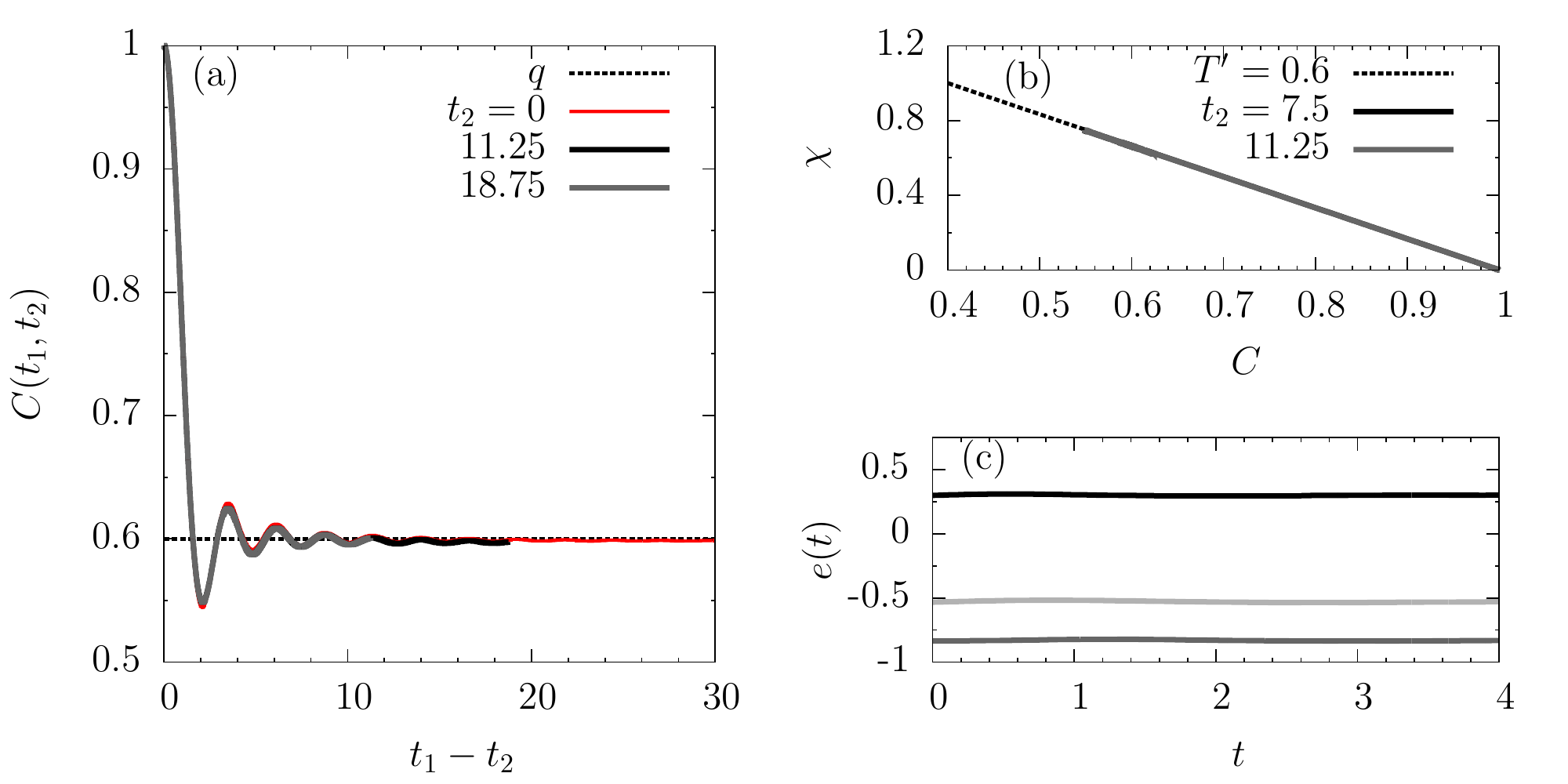}
\end{center}
\caption{\small Constant energy dynamics in the temperature region $T^0_s< T'=0.6<T^0_d$. As $J=J_0$ and $m=m_0$ there is no quench
and $\Delta e=0$. (a) Stationary dynamics of the two-time correlation
function. The asymptotic limit is  $q\neq 0$ since $T'<T^0_d$.
 (b) Linear-response vs. correlation parametric plot.
 The dashed line shows the FDT with the initial temperature.
(c) Energy time-dependence.
The data correspond to $e_{\rm kin}$ (above), $e_f$ (middle), and $e_{\rm pot}$ (below).
All the numerically evaluated values of the parameters
 $q, \, T_f=T', \, e_{\rm pot}^f$, and $e_{\rm kin}^f$
 agree (within numerical accuracy) with the ones derived analytically. See the text for more details.
}
\label{fig:T06_eq}
\end{figure}

We used two typical cases as initial states, a paramagnetic configuration and a metastable TAP state.
Figures~\ref{fig:T08_eq} and \ref{fig:T06_eq} show three plots, with the dynamics of the correlation function~(a),
the fluctuation dissipation parametric plot (b) and the
two contributions to the energy and the total energy (c), starting from equilibrium at $T'=0.8$ and $T'=0.6$, respectively.
In both cases the system is paramagnetic initially, though at $T'=0.8$ it is a proper paramagnet while at
$T'=0.6$ it is a paramagnet-looking state made of a
mixture of non-trivial metastable states, see Sec.~\ref{sec:background}.

Let us first focus on Fig.~\ref{fig:T08_eq}.
The correlation with the initial condition,
$C(t_1,0)$ (thin red line) and the ones between two different times $C(t_1,t_2)$ (grey lines) lines are identical, apart from a small deviation at short time-delays, around the
first oscillation. The two-time correlation function is invariant under time-translations, that is to say, it is a function of $t_1-t_2$ only. All the correlation functions relax to zero,
$q_0=q=0$~(a). The Lagrange multiplier (not shown) and the potential and kinetic energies (c) quickly approach
their final values and these agree with the ones predicted analytically.
The fluctuation-dissipation relation is satisfied with the temperature of the initial condition, that is the same as the one of the final state (b).
All these results are compatible with equilibrium in the paramagnetic phase.

In Fig.~\ref{fig:T06_eq} we show results for $T'=0.6$.
The Lagrange multiplier (not shown) and energy densities approach constants (c), and stationarity is satisfied as well as the FDT with the initial temperature
(b).
The main difference with the case $T'=0.8$ is that
the correlation functions, both with the initial condition and with the configuration at a waiting-time $t_2$,
relax to a non-vanishing value (a). Within numerical accuracy we observe $q_0 = q \simeq 0.6$ and this value as well as the
asymptotic potential and
kinetic energies  are consistent with the ones stemming from the analysis in Sec.~\ref{subsec:derivation}.
One can use Eq.~(\ref{eq:solq0=q}), that coincides with  Eq.~(\ref{eq:q-eq-no-quench})
and fixes the $q$ values of the non-trivial TAP states that correspond to equilibrium in the
interval $[T^0_s,T^0_d]$~\cite{KuPaVi92}, and check that
the solution for $T_f=0.6$ is $q=0.6$, the value obtained with the numerical solution of the full dynamic equations.
As regards the energy values, the kinetic energy should be $e^f_{\rm kin}=T_f/2 =0.3$ that is obtained numerically. The
potential energy is expected to be $e^f_{\rm pot} = - J^2/(2T_f) = -0.83$ which is also correct numerically. These values are added to
$e_f=T_f/2 - J^2 /(2T_f) = -0.53$, as they should.

\subsection{Energy injection}

In this subsection we explore the dynamics after energy injection ($J<J_0$ and $m=m_0$)
and we compare our results with those obtained analytically in the previous Section.

The injection of energy over an initial state with temperature $T'>T^0_d$ trivially evolves into a paramagnetic state at a temperature $T_f>T'$.
The temperature can be calculated from the conservation of energy and the fact that for the PM phase $e^f_{\rm pot}=-J^2/(2T_f)$. It is
given by Eq.~(\ref{eq:eq_temp_x}). We have checked that the numerical solution complies with these claims (not shown here).
Therefore, we shall focus on the more interesting cases with initial temperatures such that $T^0_s < T' < T^0_d$. Figure~\ref{fig:up} summarises the results
of the  concrete numerical
quenches with energy injection that we display.


\subsubsection{$T_s^0<T'<T_d^0$: from TAP to PM}

\begin{figure}[h]
\begin{center}
\includegraphics[scale= 0.6]{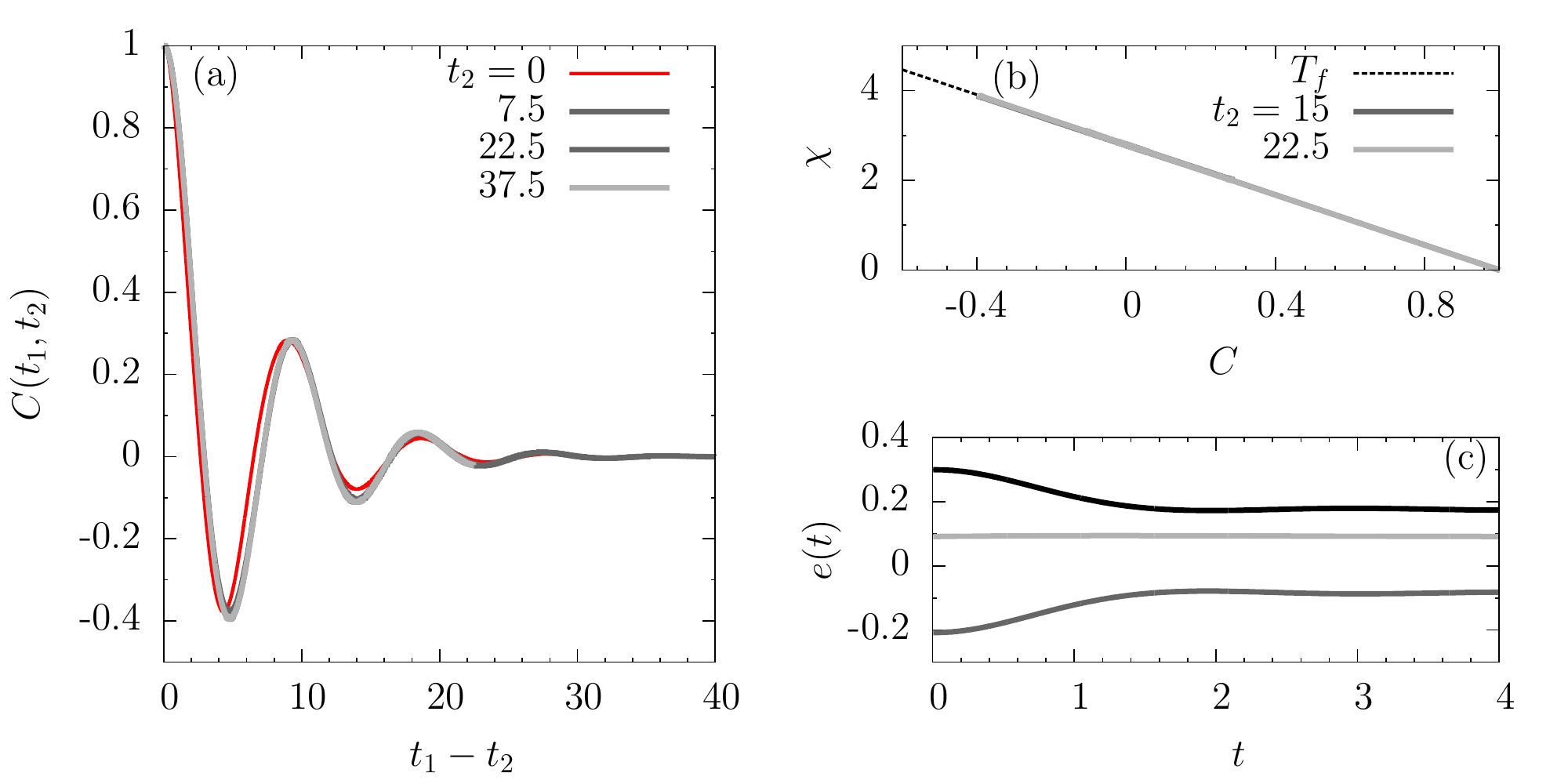}
\end{center}
\caption{\small Numerical evolution for a quench from an initial state with temperature
 in the non-trivial interval, $T^0_s < T'=0.6 < T^0_d$ and $J_0=1, \ m_0$. The final parameters are $J=0.25$ and $m=1$.
 The energy injection is large, $\Delta e=0.625$. (a) Dynamics of the correlation
function. The curves approach $q=0$ and $q_0=0$ as predicted by
Eqs.~(\ref{eq:energy-simplified2})--(\ref{eq:q-simplified}), indicating a paramagnetic equilibrium state. (b) The parametric
plot. The black dashed line shows the  FDT relation with $T_f=0.358$ as predicted by
Eq.~(\ref{eq:eq_temp_x}). The system reaches equilibrium at this new temperature.
The solid
lines correspond to the numerical results for two values of $t_2$. (c) Energy time-dependence.
From top to bottom: kinetic energy (with stationary value $e^f_{\rm kin}=0.179$),
 total energy (constant in time with value $e_f=0.092$) and potential energy (with stationary value $e^f_{\rm pot}=-0.087$).
 Note that $e^f_{\rm kin}=T_f/2$ and $e^f_{\rm pot} = -J^2/(2 T_f)$.
}
\label{fig:Tp06J025}
\end{figure}

In  Fig.~\ref{fig:Tp06J025} we show results for $T^0_s <
T'=0.6 < T^0_d$, $m=m_0$ and $J=0.25$. The system is initialised in a TAP state that corresponds to equilibrium between $T_s^0$ and $T_d^0$, see Section~\ref{subsubsec:metastability}. This quench injects a large amount of energy in the system $\Delta e=0.625$. The system is initialised in a TAP state that corresponds to equilibrium between $T_s^0$ and $T_d^0$, see Section~\ref{subsubsec:metastability}.The post-quench critical
temperatures are $T_d=0.153$ and $T_s=0.146$. The self correlations shown in (a) rapidly decay to zero for all reference times,
either when they correspond to  an initial  $t_2=0^+$ or to a waiting-time $t_2>0$. Therefore $q_0=q=0$.
These facts indicate that the system behaves as in the paramagnetic state
after the quench. The final temperature $T_f=0.358$
predicted by the asymptotic analysis, Eq.~(\ref{eq:eq_temp_x}), is in very good agreement with the numerical result extracted from the
parametric $\chi(C)$ plot in (b).
Considering the results from Section~\ref{subsec:dyn_metastable} for quenches starting from a TAP state it is important to know the temperature $T_{\mathrm{TAP}}^{\mathrm{max}}$ at which the initial TAP state ceases to exist, i.e. the position of the spinodal line in the post quench energy landscape (see~Fig.~\ref{fig:sketch-pot-energy}). For $T'=0.6$ and $J=0.25$, $T_{\mathrm{TAP}}^{\mathrm{max}}=0.186339$.
Note that $T_f>T_{\mathrm{TAP}}^{max}$ which is consistent with the system reaching a paramagnetic state with $q=q_0=0$, although it was initialised in a non-ergodic initial state, since for that final temperature the TAP state no longer exists.
From the energetic dynamics (c) we observe $e^{f}_{\rm kin}=T_f/2$ and  $e^{f}_{\rm pot}=-J^2/(2T_f)$, both results
consistent with equilibration in a paramagnetic final state. At very short times, $t_1\to 0^+$, the energies are the ones right after the
quench, $e_{\rm kin}(0^+)=T'/2$ and $e_{\rm pot}(0^+)=-JJ_0/(2T')$. It is only after a short transient, $t_{\rm tr} \simeq 1$,
that the energy densities converge to their final values.

\subsubsection{$T^0_s<T'<T^0_d$: from TAP to TAP}

\begin{figure}[h]
\begin{center}
\includegraphics[scale= 0.6]{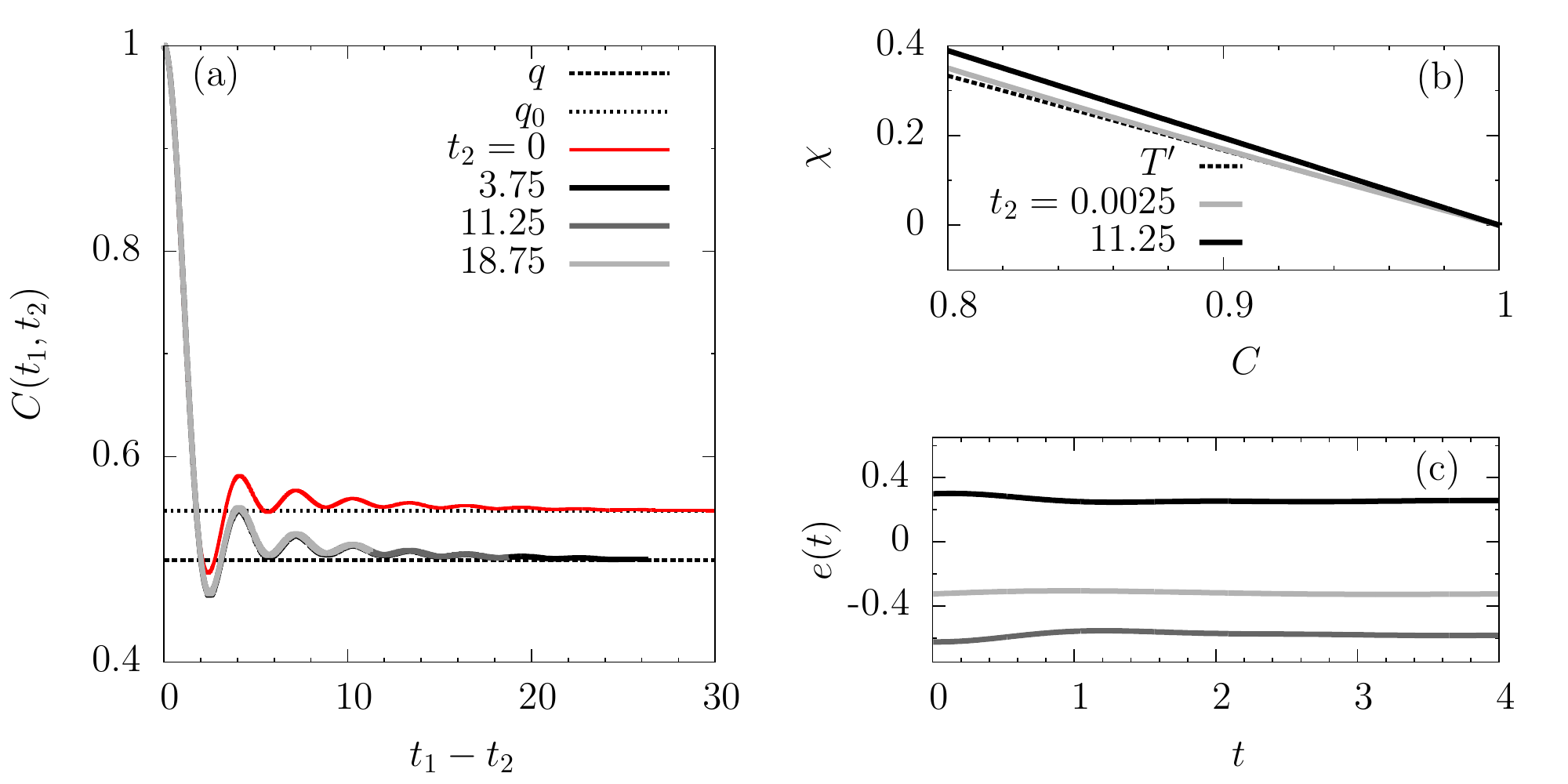}
\end{center}
\caption{\small Initial temperature in the non-trivial interval,  $T^0_s < T'=0.6 < T^0_d$, $m=m_0$ and $J=0.75$. The energy injection is small,
$\Delta e=0.208$, and the system remains trapped in a TAP state. (a) Dynamics of the correlation
function. The horizontal lines correspond to the asymptotic values $q=0.500$ and $q_0=0.548$, predicted by
Eqs.~(\ref{eq:energy-simplified2})--(\ref{eq:q-simplified}). (b) The parametric
plot. The black dotted line shows the FDT relation with $T'=0.6$ and it is compared to the numerical results
at a very early time after the quench, $t_2\simeq 0.0025$. The other numerical line was obtained with $t_2=11.25$ a sufficiently
long waiting time such that the asymptotic
$T_f=0.514$ predicted by Eqs.~(\ref{eq:energy-simplified2})--(\ref{eq:q-simplified})
is right below the data. (c) Energy time-dependence.
From top to bottom: kinetic energy (with stationary value $e^f_{\rm kin}=0.257$ and in good agreement with  $e^f_{\rm kin}=T_f/2$),
total energy (constant in time with value $e_f=0.092$) and potential energy (with stationary value $e^f_{\rm pot}=-0.582$ in agreement with Eq.~(\ref{eq:epot-finiteq})
and differently from $-J^2/(2T_f)$).
}
\label{fig:Tp06J075}
\end{figure}

In Fig.~\ref{fig:Tp06J075} we show results for $T^0_s < T'=0.6 < T^0_d$ and $J=0.75$.
This quench injects a smaller amount of energy into the system $\Delta e=0.208$. The post-quench critical temperatures are
$T_d=0.459$ and $T_s=0.439$. Differently from the previous case, the correlations with the initial time $t_2=0^+$
and with a waiting time $t_2>0$ decay to non-vanishing values, $q_0$ and $q$, respectively.
The asymptotic analysis condensed in
the full set of Eqs.~(\ref{eq:energy-simplified2})--(\ref{eq:q-simplified})
predicts  $q=0.500$ and $q_0=0.548$  that are in very good agreement
with the values obtained with the numerical solution of the dynamic equations shown in (a).

In panel (b) we display the $\chi(C)$ parametric plot for a long waiting time $t_2=11.25$, that finds good agreement with the
FDT at the final temperature $T_f=0.514$ predicted by the asymptotic analysis.
As a complement
we also plot the parametric construction for a very short waiting time,  $t_2=0.0025$, to demonstrate that,
for $C(t_1,t_2)$ very close to one, the slope is determined by the initial temperature $T'$ instead of $T_f$. It is only after a
transient that the FDT with the final temperature $T_f$ establishes.

The results in the previous paragraph are
consistent with the fact that the energies reach their asymptotic values only after a (short) transient.
From the energetic dynamics we observe that $e^{f}_{\rm kin}=T_f/2$ after $t_{\rm tr} \simeq 1$
and, $e^{f}_{\rm pot}$, measured after the same transient, is also
in very good agreement with the predictions of the asymptotic analysis, once the non-vanishing
values of $q$ and $q_0$ are taken into account.

The temperature at which the TAP state in which the system was initialised, modified by the quench,
disappears is $T_{\rm TAP}^{\rm max}=0.559$, that is slightly above the final temperature
$T_f=0.514$. Consequently, the analysis in Sec.~\ref{subsec:dyn_metastable} applies to this case and, after the quench, the
system follows the TAP state in which it was set in initially.

\subsubsection{$T^0_s<T'<T^0_d$: from TAP to spinodal, transient dynamics}
\label{subsubsec:TAP-to-spinodal}

So far we have been interested in describing the asymptotic state of the system after the quench.
We have shown that these asymptotic states can be described in terms of algebraic equations involving a few variables.
Such asymptotic equations were derived inserting appropriate \textit{Ansatze} in the full evolution equations. However,
a systematic investigation of the full dynamical equations shows that there are parameter regimes in which the system
shows long lived transient dynamics before reaching the asymptotic state. This transient effects cannot be captured by
the asymptotic analysis of the evolution equations.

We find transient dynamics near the interphases that separate the different asymptotic regimes. More precisely, near
the interphase dividing the dynamics within TAP states from the dynamics that leaves the TAP state into the PM state,
that is to say, close to the spinodal.

\begin{figure}[h]
\begin{center}
\includegraphics[scale=0.7]{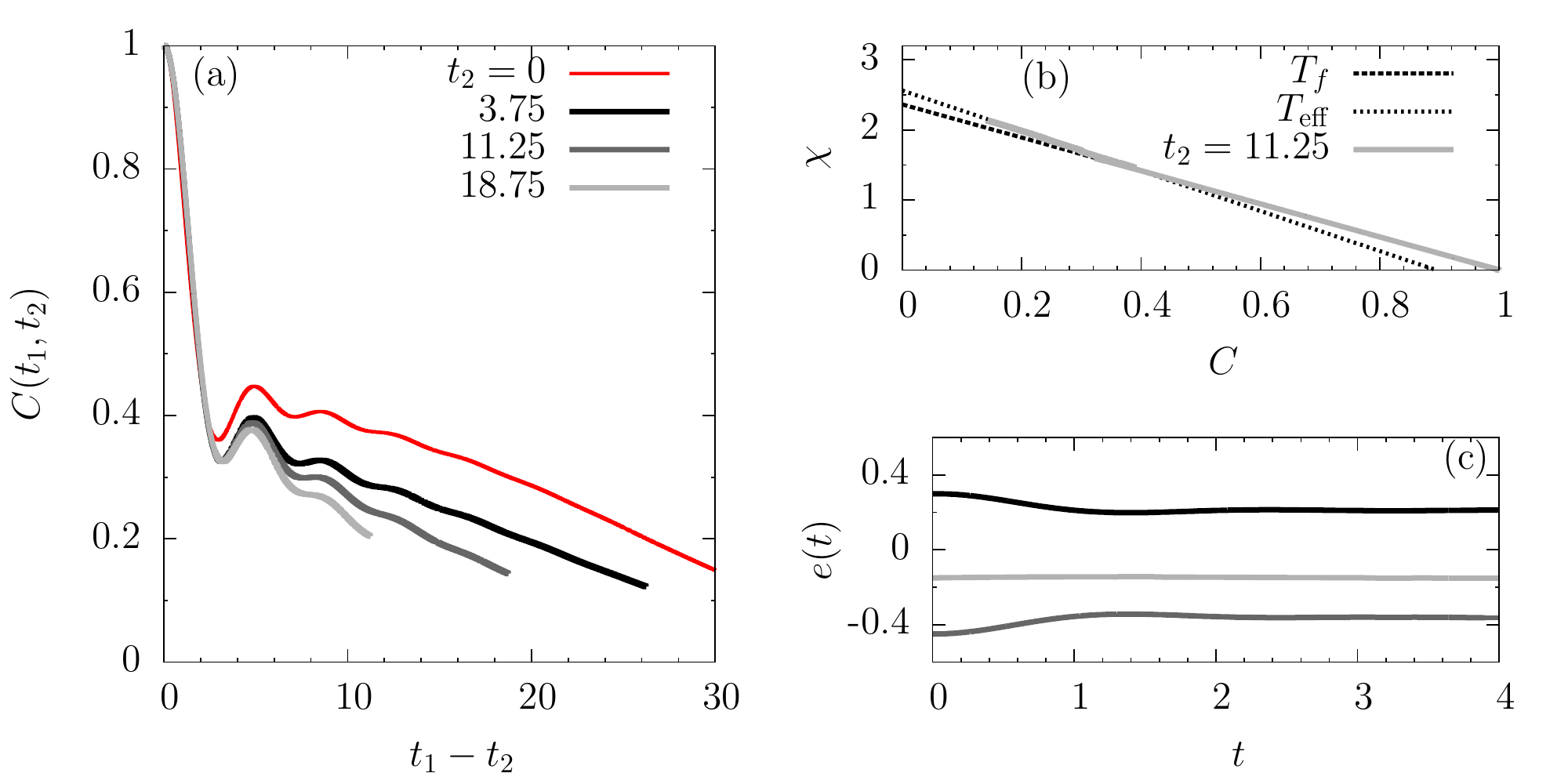}
\end{center}
\caption{\small Initial temperature in the non-trivial interval,  $T^0_s < T'=0.6 < T^0_d$, $m=m_0$ and $J=0.54$. The energy injection is
$\Delta e=0.380$. (a) Dynamics of the correlation
function. TTI is broken for these waiting times $t_2$. As the reference time increases, the correlation function relaxes more quickly,
indicating that this behaviour is a transient. (b) Parametric
plot. With solid gray line the numerical results for $t_2=11.25$. With black dashed lines the fits to the numerical data at
short and long time differences yielding $T_f \neq T_{\rm eff}$ .  (c) 
Energy time-dependence.
 From top to bottom: kinetic energy (with stationary value $e^f_{\rm kin}=0.206$), total energy (constant in time with value $e_f=-0.15$) and potential energy (with stationary value
$e^f_{\rm pot}=-0.356$).
}
\label{fig:Tp06J054}
\end{figure}

In Fig.~\ref{fig:Tp06J054} we show an example in which, starting from equilibrium below $T^0_d$, we inject energy and we observe a very long transient
regime in which the system has non-stationary dynamics, with the correlations decaying faster for longer waiting times but not yet reaching the
steady state. The non-stationary relaxation is accompanied by a waiting-time dependence of the parametric
$\chi(C)$  plot.  For short waiting times the parametric plot shows a piecewise form characteristic of ageing systems (b). However, in this case, $T_f(t_2)>T_{\rm eff}(t_2)$. In fact, a linear fit (shown with dotted lines in the figure) yields $T_f(t_2)=0.423$ and $T_{\rm eff}(t_2)=0.349$. This behaviour persists for a finite period of time before slowly approaching the asymptotic $T_f$ in the whole
range of variation of $C$ (not shown).

The asymptotic state should be paramagnetic for these parameters. Therefore, the expected $T_f$ is given by Eq.~(\ref{eq:eq_temp_x}), and
takes the value $T_f=0.41$. The predicted potential energy from the asymptotic analysis is $e^f_{\rm pot}=-0.355$, that is in very good agreement with the
numerical steady state value attained already at $t_1\simeq 2$ within numerical accuracy. The kinetic energy also reaches a
plateau after the same short transient (see the panel (c)). We note that these two ``one-time'' observables saturate much sooner that the
correlation and linear response ``two-time'' quantities converge to their final form.

The final temperature $T_f=0.41$ predicted by the asymptotic analysis is slightly above $T_{\rm TAP}^{\rm max} = 0.402$,
which justifies the PM nature of the asymptotic behaviour. However, the long-lived transient masks the PM behaviour at
not sufficiently long times.

This behaviour shares points in common with observations already made in studies in different fields. In the context of
quantum quenches,
to have non-trivial dynamics of the correlation functions while quantities such as the kinetic energy energy has already thermalised is close to the concept of prethermalisation~\cite{BerBorWe04}.
In the context of glassy physics, an asymptotic stationary decay in two steps, a faster one towards a plateau and a slower one towards zero
is the kind of relaxation found in super-cooled liquids, the hallmark of the random first order phase transition scenario. Here we
see that the correlation decays towards a value that is close to $q_{\rm TAP}^{\rm max}=(p-2)/p=1/3$, and it oscillates a few times around this
value to later decay to zero, signalling the discontinuous way in which the TAP states disappear. Finally, the inversion in the temperature
hierarchy, $T_{\rm eff}(t_2)<T_f(t_2)$, found at short waiting times $t_2$ is a transient feature that shows the memory of the
initial state with lower potential energy.
In the dissipative problem, $T_{\rm eff}>T$ for quenches from the disordered to the low temperature phase ($T<T_d$)
and $T_{\rm eff}<T$ for the dynamics in the low temperature phase for systems initiated in equilibrium at still lower
temperatures than the one at which the dynamics takes place.
This hierarchy is interpreted arguing that the effective temperature keeps memory of the initial state being more disordered
or more ordered than the target one. This behaviour has been found in numerical simulations of the out of equilibrium
dynamics of the $2d$ xy model~\cite{BeHoSe01}, an elastic line in a random potential~\cite{IgBuKoCu09} and atomic glass
models~\cite{GnMaPaSc13}, for instance. In the quenches we consider in this paper we only see the inversion in
a pre-asymptotic regime.

\subsubsection{$T^0_s<T'<T^0_d$: from TAP to threshold?}

A natural question to pose is whether it is possible to take the system out of a TAP state and put it on the
threshold level by injecting an adequate amount of energy. The asymptotic equations derived in Sec.~\ref{subsec:nonstationary}
under the assumption that this is possible allow for a non-trivial solution in a selected range of
parameters. However, for the same set of parameters the stationary state equations that describe the
dynamics within TAP states also admit non-trivial solutions. See Fig.~\ref{fig:temps} (b) where the TAP branch co-exists with the double ageing one.
The complete numerical solution  of the exact dynamic equations should then decide which of the two asymptotic
states is actually realised. We have checked this issue for, for example,
$T'=0.6$, $J_0=1$, $J=0.85$ and $m=m_0=1$, parameters such that the ageing asymptotic
solution has $q\neq 0$, $q_0=0$,  $T_f=0.542$ and $T_{\rm eff}=0.448$ while the
stationary state solutions are $q=0.55$, $q_0=0.574$ and $T_f=0.548>T_d$. The
numerical analysis of the full equations converges to the second option, showing that it is not
possible to take the system out of a TAP state and put it on the threshold.

\subsection{Energy extraction}
\label{subsec:extraction}

When extracting energy with the quench ($J>J_0$ and $m=m_0$), we will
distinguish the cases in which the initial temperature is below or above the dynamic critical temperature $T^0_d$.
We recall that in the former case the initial configuration is drawn from a non-trivial TAP state while in the latter it is
simply paramagnetic. By extracting a small amount of energy from a highly energetic PM state the system remains
in the PM state; these cases are not particularly interesting and we do not show any example of such. Instead, we
focus on more interesting cases that are summarised in Fig.~\ref{fig:down}.

\subsubsection{$T^0_s<T'<T^0_d$: from TAP to TAP}
\label{subsubsec:lowTprima}

\begin{figure}[h]
\begin{center}
\includegraphics[scale= 0.6]{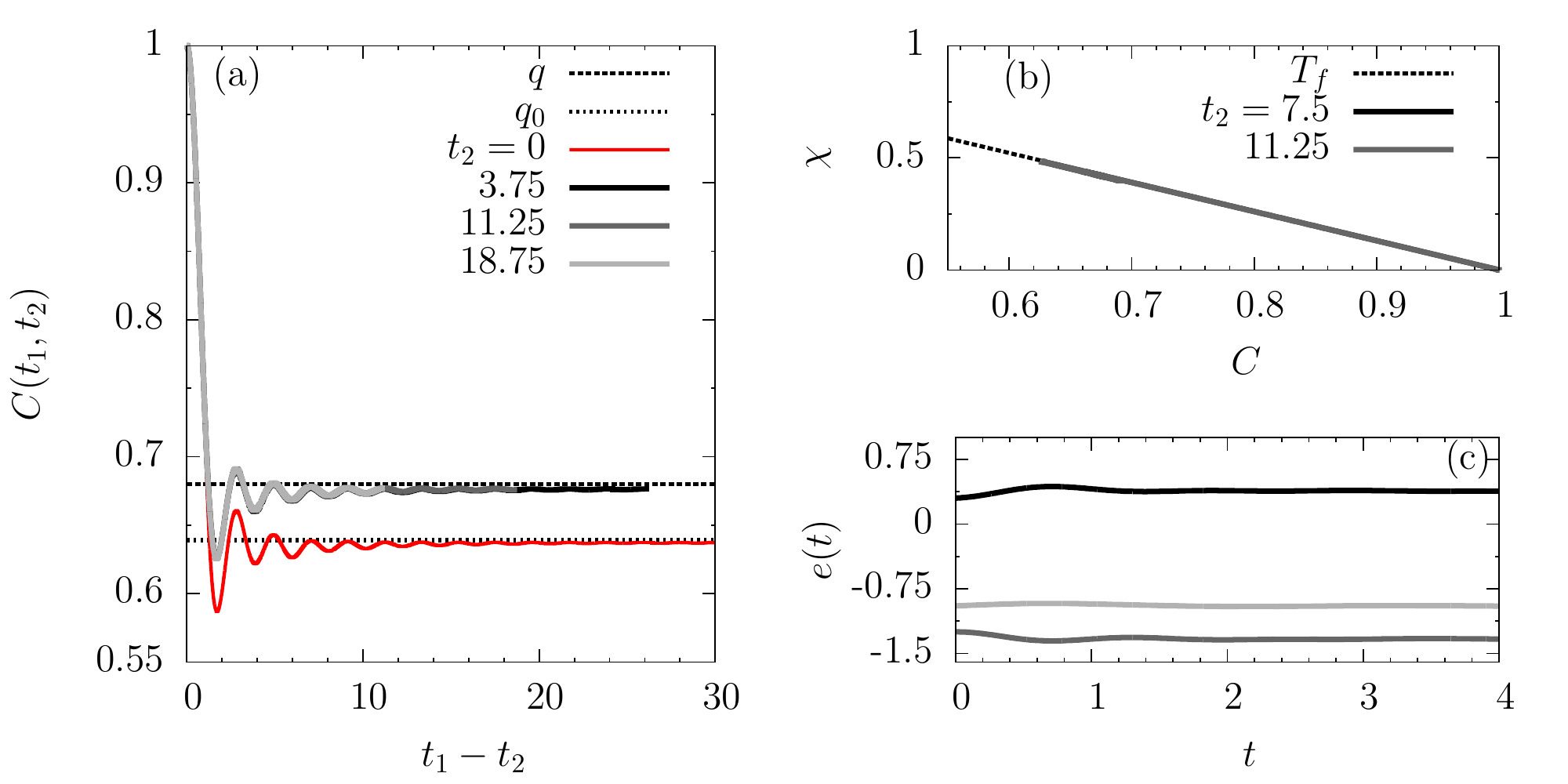}
\end{center}
\caption{\small Starting from $T^0_s < T'=0.6 < T^0_d$, $m=m_0$ and we use $J=1.5$. The energy extraction is
$\Delta e=-0.416$. (a) Dynamics of the correlation
function. The horizontal dotted lines correspond to the asymptotic values $q=0.680$ and $q_0=0.639$,
predicted by Eqs.~(\ref{eq:energy-simplified2})--(\ref{eq:q-simplified}). (b) Parametric
$\chi(C)$ plot.
The
solid lines are the numerical results, in  perfect agreement with the analytic prediction for the FDT
with $T_f=0.766$,  shown with a black dashed line.  (c)
Energy time-dependence.
From top to bottom: kinetic energy (with stationary value $e^f_{\rm kin}=0.383 =T_f/2$), total energy (constant in time with value $e^f=-0.950$) and
potential energy (with stationary value $e^f_{\rm pot}=-1.333$).
}
\label{fig:Tp06J15}
\end{figure}

In Fig.~\ref{fig:Tp06J15} we show results for $T^0_s < T'=0.6 < T^0_d$ and $J=1.5$. This quench extracts a large amount of energy
from the system $\Delta e=-0.416$.
For such value of $J$ the critical temperatures are $T_d=0.918$ and $T_s=0.440$. The final temperature
$T_f=0.766$ and the parameters $q=0.680$ and $q_0=0.639$ predicted by the asymptotic analysis are in very good agreement with the
results from the numerical solution of the complete equations.
In this case $T_s<T_f<T_d$. From the energetic evolution we
observe that $e^{f}_{\rm kin}=T_f/2$. In parallel, the potential energy
$e^{f}_{\rm pot}$ is also in very good agreement with the predictions of the asymptotic analysis
using the non-vanishing values of $q$ and $q_0$.

This case is an example in which the initial TAP state is followed by the dynamics. The description
in Sec.~\ref{subsec:dyn_metastable} applies and explains the results.

\subsubsection{$T'>T^0_d$: from PM to threshold, ageing dynamics}
\label{subsubsec:highTprima}

We will now demonstrate that for quenches  from the  paramagnetic state, $T'>T_d$, with  sufficient extraction of energy
the system approaches the threshold level, similarly to what has been been observed in the past for the relaxation dynamics of the model coupled to a
thermal bath. Due to the flatness of this region of phase space we observe ageing phenomena with violations of the fluctuation dissipation
theorem and the appearance of an effective temperature even with conserved energy dynamics.

\begin{figure}[h]
\begin{center}
\includegraphics[scale=0.6]{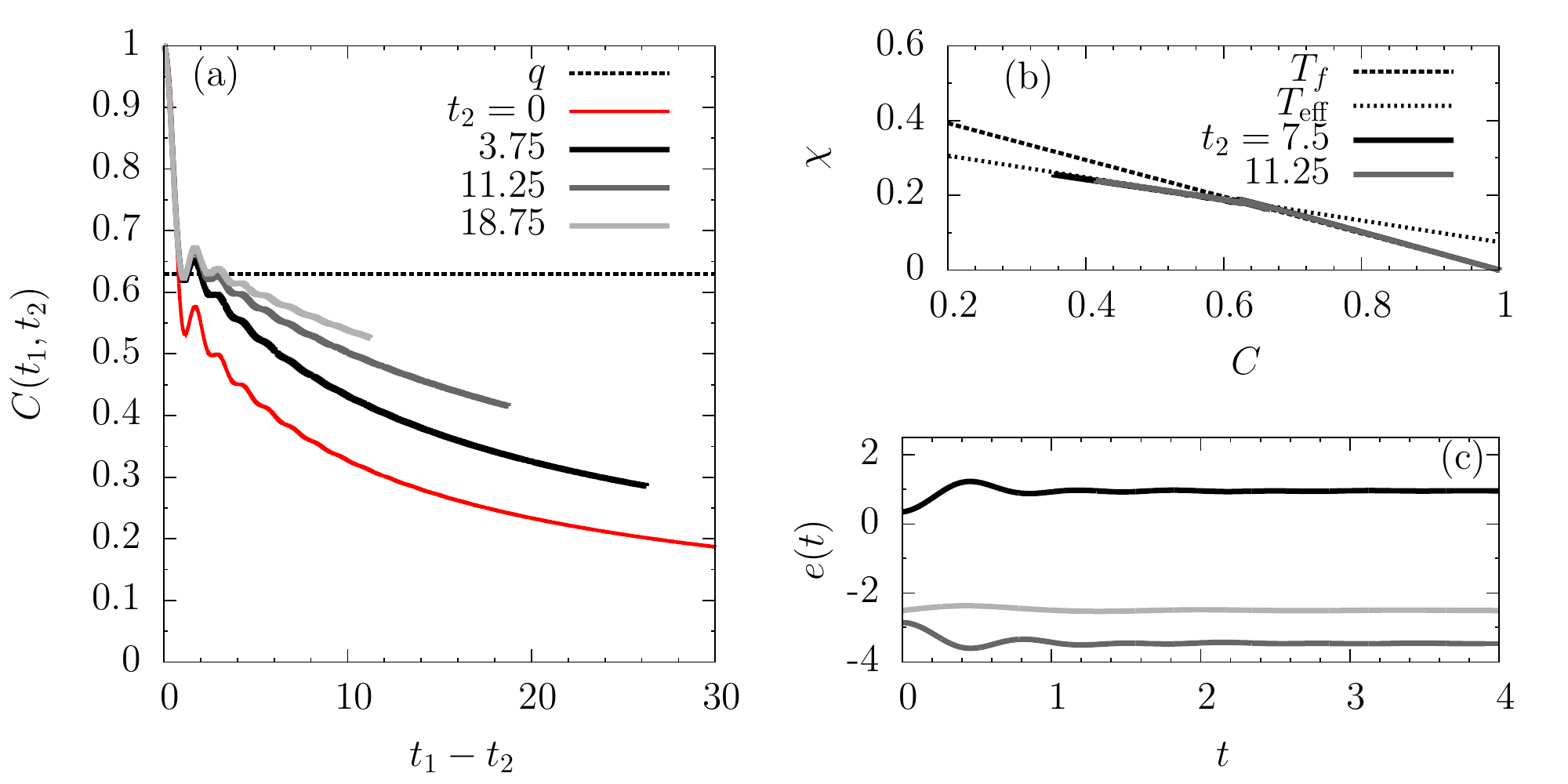}
\end{center}
\caption{\small
Starting from $T'=0.7 > T^0_d$ and $m_0$ we choose $m=m_0$ and $J=4$. The energy extraction is
$\Delta e=-2.175$. (a) Dynamics of the correlation
function. TTI is broken and ageing is evident. The dashed horizontal line corresponds to the value of $q$ predicted by the asymptotic equations,
Eqs.~(\ref{eq:q-const-energy}),~(\ref{eq:Teff-const-energy}) and~(\ref{eq:ef-const-energy}). (b) Parametric
plot. Black dashed and dotted lines are the predictions from the asymptotic equations with $T_f=2.036$ and $T_{\rm eff}=3.461$.
We also show numerical results for two different waiting times that are indistinguishable on the plot. (c)
Energy time-dependence.
From top to bottom:
kinetic energy (with stationary value $e^f_{\rm kin}=0.995$), total energy (constant in time with value $e_f=-2.505$) and potential energy (with stationary value
$e^f_{\rm pot}=-3.501$).
}
\label{fig:Tp07J4}
\end{figure}

In Fig.~\ref{fig:Tp07J4} we show results for energy extraction starting from a paramagnetic state, $T'>T^0_d$, and using $J=4$, a value for which
$T_d = 2.324$ and $T_s = 2.449$.

It is clear from panel (a) that the correlation function does not reach a stationary regime;
hence, time-translational invariance is broken. Moreover,  the system ages since the curves for longer waiting times decay in a slower manner than the ones
for shorter waiting times. The correlation shows oscillations at small values of the time-delay and these
progressively disappear at long values of the same time-delay. The decay of any of the curves for different waiting times,
but especially the ones for long waiting time, occurs in two steps.

The parametric plot $\chi(C)$ in Fig.~\ref{fig:Tp07J4} (b) does not show a waiting-time dependence, as proven
by the fact that the curves for two values of $t_2$ fall on top
of each other. The resulting master curve also has a two step structure, with two slopes, which are in very good agreement with the results from the asymptotic equations Eqs.~(\ref{eq:q-const-energy}),~(\ref{eq:Teff-const-energy}) and~(\ref{eq:ef-const-energy}) for $T_f$ and $T_{\rm eff}$.
The breaking point in the
piece-wise straight line is at $C\simeq 0.6$,
which agrees with the value of $q$ predicted by the asymptotic equations, $q=0.629$, and the value of the change in behaviour of the two-time correlation,
see panel~(a).

Panel (c) shows the evolution of the two contributions to the energy density.
We verify that the asymptotic kinetic energy density in Fig.~\ref{fig:Tp07J4} (c) is consistent with $T_{f}/2$.
Moreover, the stationary potential energy density found numerically is also in very good agreement with the prediction from
Eq.~(\ref{eq:pot_aging}), $e^f_{\rm{pot}}=e_{\rm th} = -3.525$.

For intermediate energy extraction the system explores regions near the threshold level but still in the PM part of the landscape. As a consequence,
there appear transient regimes in the dynamics at short times with non-stationary correlations that resemble the ageing ones,
and a $\chi$ vs. curve that can be characterised with two temperatures. However, for longer waiting times the dynamics converge to the
asymptotic PM solution.

\subsection{Summary}

\begin{figure}[h]
    \begin{minipage}[c]{1.00\linewidth}
        \begin{center}
        \includegraphics[scale=0.7]{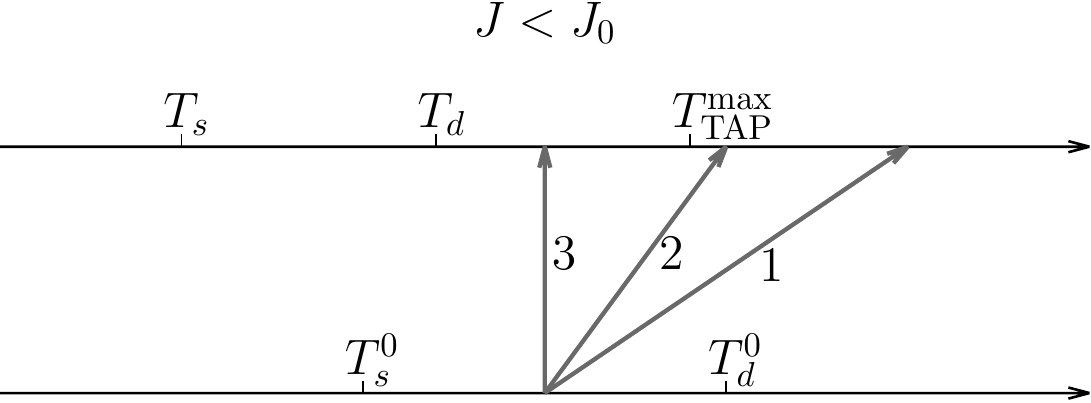}
        \end{center}
    \end{minipage}%

    \begin{minipage}[c]{1.00\linewidth}
        \begin{center}
        \begin{tabular}[c]{|l|c|c|c|c|c|c|l|}
            \hline
            & $J$ & $\Delta e$ & $T_d$ & $T_f$ &  $T_{\rm eq}^{\rm max}$ &  $T_{\rm TAP}^{\rm max}$ & \mbox{Asymptotic state} \\
            \hline
            \hline
            \mbox{1 (Fig.~\ref{fig:Tp06J025}}) & 0.25 & 0.625 &0.153 & 0.358 & 0.198 & 0.186 & \mbox{PM} \\
            \hline
            \mbox{2 (Fig.~\ref{fig:Tp06J054}}) & 0.54 & 0.380 & 0.330 & 0.423 & 0.427 & 0.402 \, & \mbox{slow approach towards PM} \\
            \hline
            \mbox{3 (Fig.~\ref{fig:Tp06J075}}) & 0.75 & 0.208 & 0.459 & 0.518 & 0.593 & 0.559 & \mbox{TAP}
            \\
            \hline
        \end{tabular}
        \end{center}
    \end{minipage}

\caption{\small Schematic representation of the initial and final states of the quenches with energy injection studied numerically in Figs.~\ref{fig:Tp06J025}
and \ref{fig:Tp06J075}. The numbered labels refer to the entries in the table, in which we show quantitative information for each illustrative case.
}
\label{fig:up}
\end{figure}

The results of the energy injection process are summarised in the Table included in Fig.~\ref{fig:up}. The simplest way to understand what is going on
is to compare the final temperature $T_f$ to the characteristic temperatures
after the quench.  In the three cases shown, $T_f>T_d$. However, the comparison between $T_f$ and $T_{\rm TAP}^{\rm max}$ that
corresponds to the spinodal line (see Fig.~\ref{fig:sketch-pot-energy}) explains the different
behaviour in the three quenches. For $J=0.25$, $T_f>T_{\rm TAP}^{\rm max}$ and the only possibility is to have a paramagnetic
behaviour, as seen in Fig.~\ref{fig:Tp06J025}. For $J=0.54$,
$T_f \stackrel{>}{\sim} T_{\rm TAP}^{\rm max}$ for the TAP state in which the system was initialised and, therefore, the system needs a long time to
relax to the PM solution, see Fig.~\ref{fig:Tp06J054}. Finally, for $J=0.75$ the TAP state still exists after the quench and the system just
follows it, as explained in Sec.~\ref{subsec:dyn_metastable}, see Fig.~\ref{fig:Tp06J075}.

\begin{figure}[h]
    \begin{minipage}[c]{1.00\linewidth}
        \begin{center}
        \includegraphics[scale=0.7]{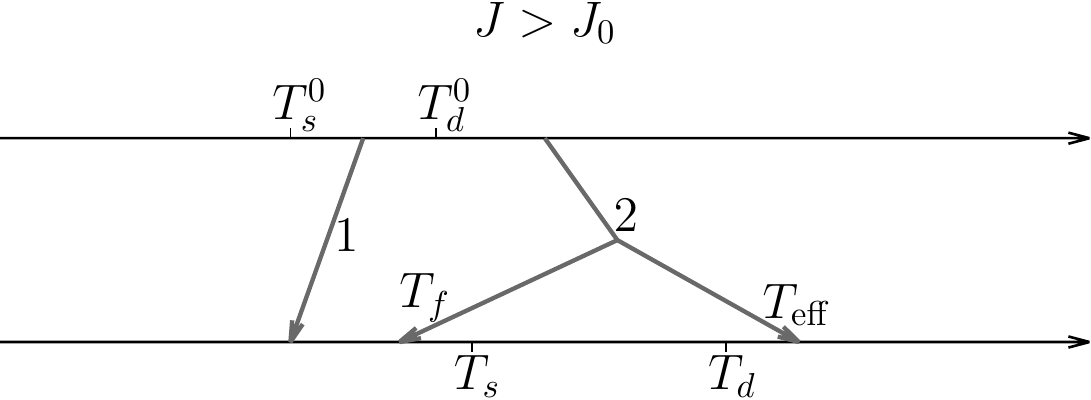}
        \end{center}
    \end{minipage}

    \begin{minipage}[c]{1.00\linewidth}
        \begin{center}
        \begin{tabular}{|l|c|c|c|c|c|c|l|}
            \hline
            & $J$ & $\Delta e$ & $T_d$ & $T_f$ &  $T_{\rm eff}$ &  $T_{\rm TAP}^{\rm max}$ & \mbox{Asymptotic state} \\
            \hline
            \hline
            \mbox{1 (Fig.~\ref{fig:Tp06J15}}) & 1.5 & $-0.416$  & 0.918 & 0.766 & - & 1.118 &  \mbox{TAP} \\
            \hline
            \mbox{2 (Fig.~\ref{fig:Tp07J4}}) & 4 &  $-2.175$ & 2.448 & 2.036  & 3.461 & - &  \mbox{Threshold} \\
            \hline
        \end{tabular}
        \end{center}
    \end{minipage}

\caption{\small Schematic representation of the initial and final states of the quenches with energy extraction studied numerically in Figs.~\ref{fig:Tp06J15} and \ref{fig:Tp07J4}.
The numbered labels make reference to the entries of the table, in which we show quantitative information for each illustrative case.
}
\label{fig:down}
\end{figure}

We recap the two observations made for the energy extraction process  in Fig.~\ref{fig:down}. The distinction is due to the initial state,
being above or below $T_d$. In the latter case the system can only follow the TAP state in which it was prepared. In the former
the parameters can be tuned to set it on the threshold.

\section{The phase diagram}
\label{sec:phase-diagram}

\begin{figure}[h]
\begin{subfigure}{\linewidth}
\begin{center}
\includegraphics[scale=0.65]{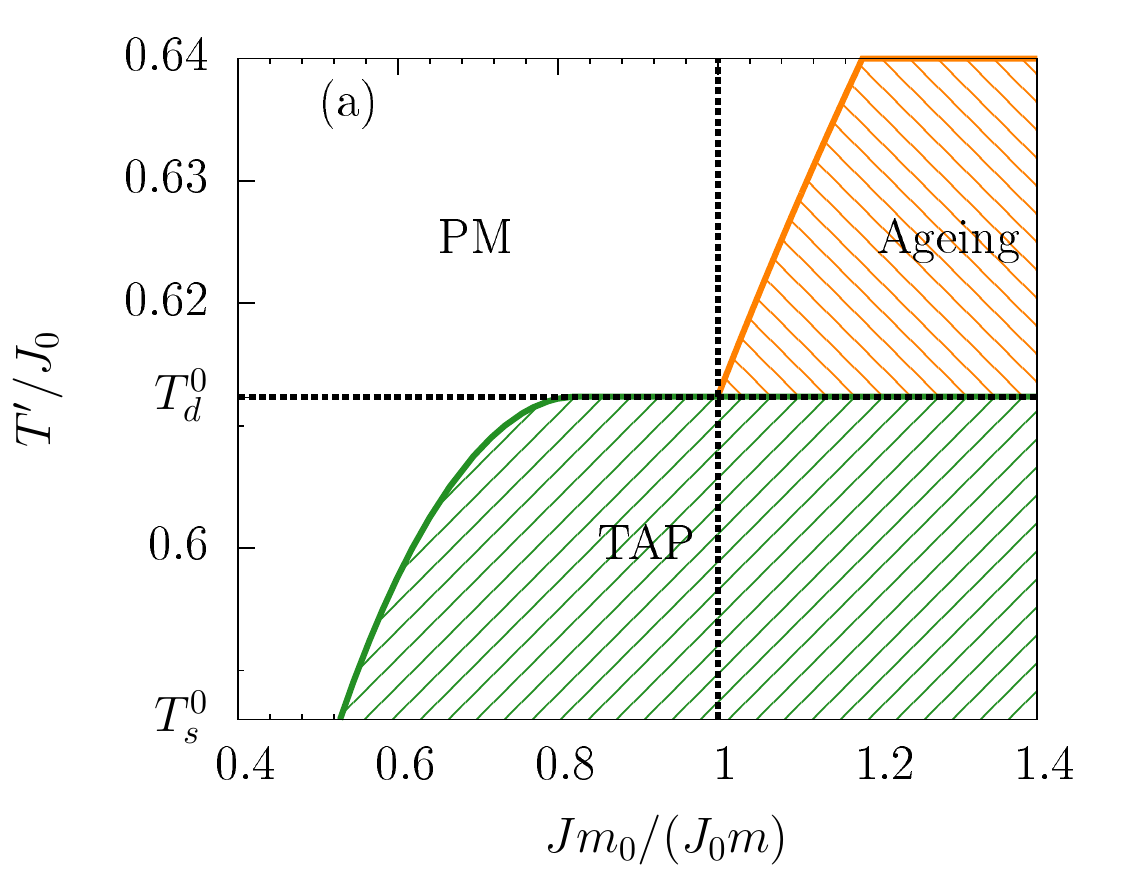}
\includegraphics[scale=0.65]{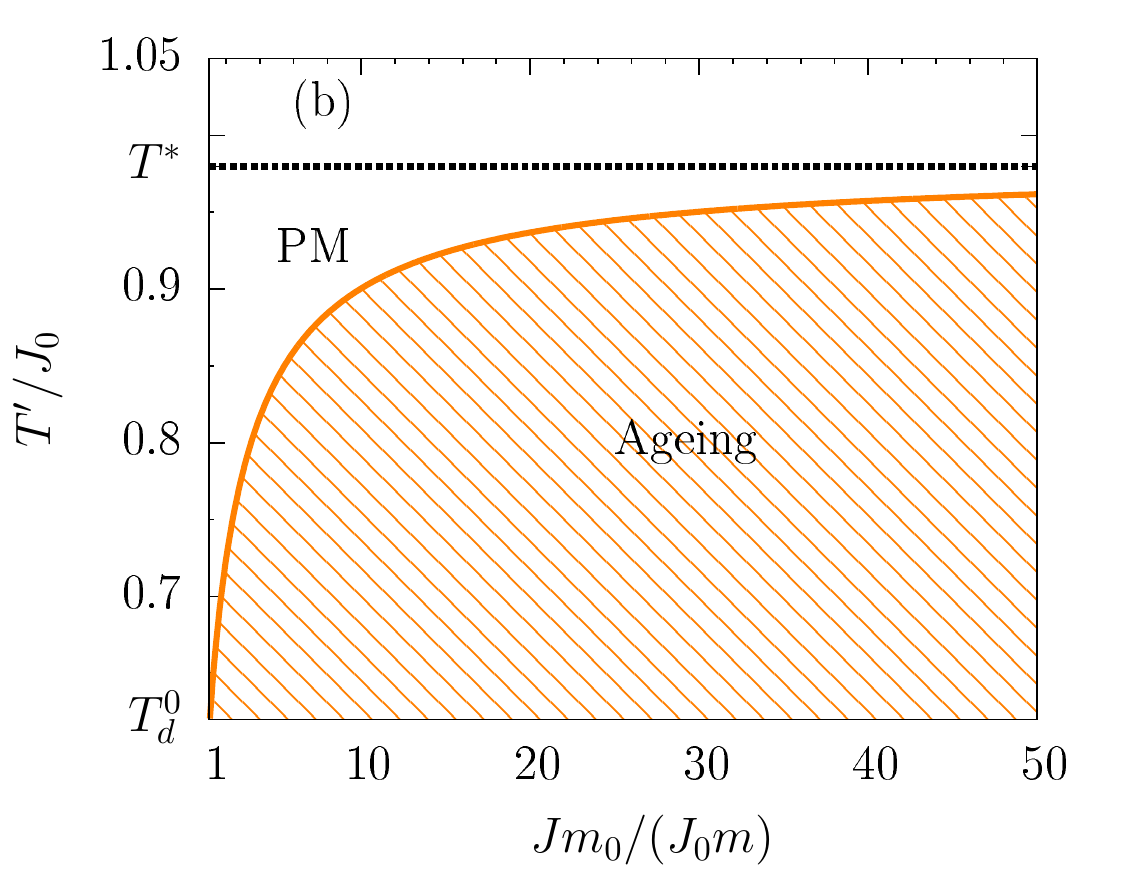}
\end{center}
\end{subfigure}
\caption{\small The phase diagram $(Jm_0/(J_0m), T'/J_0)$ with $T'$ the temperature of the initial condition
$J_0$ and $J$ the parameters that characterise the width of the disorder distribution before and after the quench,
and $m_0$ and $m$ the mass of the particle before and after the quench. $Jm_0/(J_0m)>1$ represents energy
extraction and $Jm_0/(J_0m)<1$ energy injection.
In (a) we show the region around $T'=T_d^0$ (low temperatures), while in (b) we show the region $T'>T_d^0$.
The analytic expressions for the boundary lines are obtained in this Section.
}
\label{fig:phase-diagram}
\end{figure}

The purpose of this Section is to determine a dynamical phase diagram.
We choose as the vertical axis  the temperature at which the initial condition was equilibrated
normalised by the parameter $J_0$. The horizontal axis is the control parameter of the
quench that, as we will show, turns out to be $Jm_0/(J_0m)$.
We will show the regions in which the {\em final} state is either paramagnetic, a TAP state or
a non equilibrium ageing state.
As in the rest of this paper we show results for initial states equilibrated at temperatures  $T_s^0<T'$ only, so the origin of this axis
is at $T_s^0$.
In the rest of the Section we explain the criteria used to obtain the critical lines
in Fig.~\ref{fig:phase-diagram}.

\subsection{The PM-ageing boundary line}

 Initial states with  $T'>T_d^0$ are paramagnetic. We have seen in Sec.~\ref{subsec:extraction} that extracting a small amount of energy, leaves the state in the
 paramagnetic region. Instead, by extracting a larger amount of energy it is possible to put the system on the threshold and, accordingly,
 the system displays ageing dynamics and is  characterised
 by $T_{\rm eff}>T_f$. If we claim that this second option ceases to be possible when
 $T_{\rm eff}=T_f$, we can then use Eq.~(\ref{eq:Teff-const-energy})
to derive the value of $q$ on the transition line.
\begin{equation}
1 = (p-2)(1-q_{\rm cr})/q_{\rm cr} \qquad\Rightarrow\qquad q_{\rm cr} = \frac{p-2}{p-1}
\; ,
\end{equation}
that is $q_{\rm th}(T_d)$.
If we now replace this $q_{\rm cr}$ in Eq.~(\ref{eq:q-const-energy}) and we use the constant $a(p)$, see Eq.~(\ref{eq:adep}),
\begin{equation}
{T_f^2}_{\rm cr}  = J^2 a^2(p)  = T^2_d
\; ,
\end{equation}
independently of $T'$ and $J_0$, see Fig.~\ref{fig:temps}. Still, what we are looking for is the curve
$T' (J_0,J)$ on which $T_f$ takes this value. We   obtain it from Eq.~(\ref{eq:ef-const-energy}) evaluated at $T_f = T_{\rm eff} = T_d$
and $q=q_{\rm cr}$
\begin{eqnarray}
\frac{T'}{J_0} = \frac{1}{2} \frac{1}{a(p)} \frac{Jm_0}{J_0 m}
\;
\left[
a^2(p) - 1
+ \sqrt{\left(a^2(p)- 1 \right)^2 + 4 \, a^2(p) \, \frac{J_0m}{Jm_0}}
\right]
\; .
\label{eq:a}
\end{eqnarray}
A Taylor expansion around $Jm_0/(J_0 m) =1$ yields, to first order,
\begin{equation}
\frac{T'}{J_0} = a(p) + \left(\frac{Jm_0}{J_0m}-1 \right) \frac{a(p)}{1+a^2(p)}
\end{equation}
and, in the particular case $Jm_0/(J_0 m ) =1$,
\begin{equation}
T' =  a(p) J_0 = T_d^0
\; .
\end{equation}
In the limit $Jm_0/(m J_0) \to\infty$, using $a^2(p)<1$, Eq.~(\ref{eq:a}) implies
\begin{eqnarray}
\frac{T'}{J_0} \to \frac{T^*}{J_0} \equiv \frac{a(p)}{1-a^2(p)}
\; ,
\label{eq:aa}
\end{eqnarray}
a finite value for all $p$; in particular,
$T'/J_0\to 2\sqrt{6}/5$ for $p=3$.
This seemingly unexpected result can be rationalised as follows.
For increasing $T'$, the initial kinetic energy (at $t_1=0^+$) grows as $T'$, while the initial potential energy vanishes as $-J_0J/(2T')$ and the energy extraction, in the
case in which, for concreteness, we apply a quench in the potential, as $J_0(J_0-J)/(2T')$. In the final ageing state the two temperatures
$T_f$ and $T_{\rm eff}$
are finite for finite $J$, as one can simply verify from the asymptotic equations. Accordingly, for finite $J$
there is a maximal value of $T'$ beyond which the initial kinetic energy cannot be extracted by the chosen $J$ to put the system on the threshold.
The paramagnetic solution does not have this problem since its $T_f$ is not bounded in the same way
for quenches with $T'/J_0 > T^*/J_0$. The system then remains in the
PM state.
Contrary to this limitation, in the thermal quenches of the over-damped dissipative model, the system can get rid of its extra energy by releasing it to the environment
and quenches from arbitrary high temperature initial conditions can approach the threshold and show ageing. Another interesting
feature is that the final $T_f$ cannot take arbitrary low values.

\subsection{The TAP-PM boundary line}

Suppose now that we start in a TAP initial state. As we have seen, if we inject a small amount of energy, the
system finishes in the same TAP state of the post-quench potential. Nevertheless, if we inject a larger amount of energy, the system can end
up in a paramagnetic final state.
The TAP-PM  boundary line  should be determined by the impossibility to follow the TAP initial state at the target $J$.
Therefore, the transition occurs on the spinodal, where the TAP states simply cease to exist (see the blue line and open squares
in Fig.~\ref{fig:temps}~(b)).

The critical line $(Jm_0/(J_0m), T'/J_0)$ can then be derived by exploiting the results in Sec.~\ref{subsec:dyn_metastable}.
First, Eq.~(\ref{eq:relation_if}) relates the initial $q[J_0, T']$  to the final one $q[J,T_f]$. We can use the, by now usual,
analysis of the bell-shaped r.h.s. to deduce that a solution with  $q[J,T_f]\neq 0$ exists as long as
$q \leq q_{\rm max}= (p-2)/p$. Evaluating then the r.h.s. at this value
\begin{equation}
1- q[J_0,T'] = \frac{2 J^2}{p T^2_f} \; \left( \frac{p-2}{p}\right)^{p-2}
\; ,
\label{eq:qout}
\end{equation}
with $q[J_0,T']$ given by Eq. (\ref{eq:initial_q}) that we recall here written in a more convenient way
\begin{equation}
\frac{p}{2} (q[J_0,T'])^{p-2} (1-q[J_0,T']) = \frac{{T'}^2}{J_0^2}
\; .
\label{eq:qin}
\end{equation}

We now need an equation to fix $T_f$. This should be derived from the asymptotic
dynamics part, exploiting the fact that $q[J,T_f] = q_{\rm max} = (p-2)/p$. Take Eq. (\ref{eq:energy-simplified2}) as a starting point. The
only unknown (apart from $T_f$) is $q^p_0$. We can use the energy balance Eq.~(\ref{eq:q-simplified}) to extract $q_0^p$ and
get from (\ref{eq:energy-simplified2})
\begin{equation}
\frac{J_0m}{Jm_0} \frac{T'}{J_0} - \frac{J_0}{T'} =
\frac{T_f}{J} - \frac{J}{T_f} - \frac{2}{p} \frac{q_{\rm max}}{1-q_{\rm max}} \frac{T_f}{J}
+ \frac{J}{T_f} q_{\rm max}^{p-1}
\; .
\end{equation}
This expression simplifies a little bit replacing $q_{\rm max} = (p-2)/p$ and $q_{\rm max}/(1-q_{\rm max}) = (p-2)/2$,
\begin{equation}
\frac{J_0m}{Jm_0} \frac{T'}{J_0} - \frac{J_0}{T'} =
\frac{2}{p} \; \frac{T_f}{J} - \frac{J}{T_f}  \left[ 1 - \left(\frac{p-2}{p}\right)^{p-1} \right]
\; ,
\label{eq:ll}
\end{equation}
again a quadratic equation for $T_f/J$. Now we have to replace the solution for $T_f/J$ in Eq.~(\ref{eq:qout}),
use this linear equation on $q[J_0,T']$ to get its dependence on the parameters and
replace it in Eq.~(\ref{eq:qin}). This is an implicit equation that yields the curve $(J_0m/(Jm_0),T'/J_0)$ that
marks the end of the TAP region of the phase diagram for $T_s<T'<T_d^0$.

We can see whether this boundary touches the value $T'=T^0_d$ at the quench parameter
$Jm_0/(J_0m)=1$ or elsewhere by setting $T'=T_d^0=a(p) J_0$, with initial $q$ value  $q[J_0,T']= (p-2)/(p-1)$. Equation~({\ref{eq:qout})
can be used to determine $T_f$:
\begin{equation}
\frac{T^2_f}{J^2} = 2\; \frac{p-1}{p} \left( \frac{p-2}{p} \right)^{p-2} = 4 \left(\frac{p-1}{p}\right)^{p} a^2(p)
\end{equation}
that replaced in Eq.~(\ref{eq:ll}) yields the critical value of $Jm_0/(J_0m)$:
\begin{equation}
\left(\frac{Jm_0}{J_0m}\right)_{\rm cr}
=
a^2(p)
\;
 \left\{ 1+
\frac{4}{p} \left( \frac{p-1}{p} \right)^{p/2} a^2(p)
- \frac{1}{2} \left( \frac{p}{p-1} \right)^{p/2} \left[
1- \left(\frac{p-2}{p}\right)^{p-1}
\right]
\right\}^{-1}
\; .
\end{equation}
In the case $p=3$ one has $a^2(3) = 3/8$ and
\begin{equation}
\left(\frac{Jm_0}{J_0m} \right)_{\rm cr}
\approx 0.82
\; ,
\end{equation}
that is smaller than one, as shown in Fig.~\ref{fig:phase-diagram}.
This value of $\left(\frac{Jm_0}{J_0m} \right)_{\rm cr}$ provides the minimum energy injection needed to reach temperatures above the limit of existence of the metastable states close to the threshold level, see Figure~\ref{fig:sketch-pot-energy}.

\section{Conclusions}
\label{sec:conclusions}

We studied the dynamical evolution of a classical disordered model subject to a quench. By endowing the system with an
intrinsic dynamics, we were able to investigate the evolution of the {\em isolated} model, thus analysing issues of thermalisation and
equilibration in a classical setting.

We showed that, depending on the parameters used in the instantaneous quench,
an {\it interacting classical  disordered model} can either reach equilibrium or remain out of equilibrium
in two different manners: it can be confined in a  metastable state evolving with
stationary dynamics characterised by a single temperature related to the final energy, or it can
evolve in the so-called threshold level with non-stationary dynamics characterised by
two temperatures, similarly to what happens in the dissipative model.
Figure~\ref{fig:phase-diagram} summarises the different phases and transition lines
in the phase diagram parametrized by the most important parameters of the initial condition ($T'/J_0$)
and the adimensional control parameter of the quench, $Jm_0/(J_0m)$.
We therefore showed that dynamic phase transitions in close interacting systems are also realised classically.


\vspace{0.25cm}

In the context of quantum quenches, an out of equilibrium dynamic transition was found in the Hubbard
model~\cite{EcKoWe09,ScFa10,TsEcWe13,TsWe13}, the Bose-Hubbard model~\cite{ScBi10}
and the O(N) model~\cite{GACa11,ScBi11,ScBi13,MaChMiGa15}. It is characterised
by the fact that long time averages display a singular behaviour and the order parameter vanishes when post-quench coupling parameter, say
$U_f$, approaches a critical value $U^d_f$ with  critical dynamics~\cite{ScBi13}. On one and another side of the critical
parameter the asymptotic values of these time-averages take qualitatively different behaviours, as in a conventional
phase transition. The dynamic phase transition can be a feature of the pre-thermalisation regime, as in~\cite{ScBi11}, or
a fully asymptotic property as in the $O(N)$ model in the large $N$ limit~\cite{MaChMiGa15}.

Compared to these problems,
 in this paper we showed that dynamic phase transitions in quenched close interacting systems are also realised classically.
 The dynamical phase diagram in Fig.~\ref{fig:phase-diagram} summarises the different phases and transition lines.

\vspace{0.25cm}

Let us now explain, in some detail, why we claim that the non-integrable model that we analysed
shows equilibrium and out of equilibrium asymptotic dynamics.
In the case of a closed system, the process of thermalisation
consists in the loss of memory of the details of the initial condition, and the
approach to a state that can be completely characterised by the values of the conserved quantities.
Equilibrium statistical mechanics indicates that such a state should be described by the microcanonical, canonical or grand-canonical
ensembles  in the thermodynamic limit, depending on the degree of isolation from the environment.
Then, at least for generic non-integrable models, we can say that a closed system has thermalised if
\begin{itemize}
\item the dynamics has reached a stationary state,
\item this state is described by one of the ensembles of statistical mechanics.
\end{itemize}
Assuming thermalisation, the temperature of the canonical distribution
describing the asymptotic state of a system with constant particle number is univocally defined by its energy.
More specifically, if $H$ is the Hamiltonian of the system and $\rho_0$ is the initial macrostate, the energy of the system is given by $E=\langle H \rangle_0$, where the subscript indicates that we are taking the statistical average with respect to $\rho_0$. The final canonical state is then $\rho_{\rm th}=e^{-\beta H}/Z$, with $\beta$ determined by $E=\langle H\rangle_{\rm th}$, where the subscript indicates the thermal average.

In the model that we studied, we found that, depending on the quench parameters, the asymptotic dynamics occur on the threshold, within a TAP
metastable state or in the paramagnetic state. We now explain why the first two cases are out of equilibrium while the latter is in equilibrium, according to the
two conditions listed above.

In the parameter regimes in which the ageing solution is realised, it is quite clear that the system does not thermalise, simply because it never reaches a stationary state.

We now argue that whenever the dynamics of the system is confined in a TAP state, i.e. $q,q_0\neq 0$, although the asymptotic state is stationary
and even satisfies FDT with respect to its asymptotic temperature $T_f$, the system does not thermalise in the sense discussed above.
Let us consider two different initial conditions specified by the values of $J_0$ and $T'$,
$(1)=\{J^{(1)}_0,\ T'^{(1)}=0.6 \, J^{(1)}_0\}$ and $(2)=\{J^{(2)}_0=1.05 \, J^{(1)}_0, \ T'^{(1)}=0.62 \, J^{(1)}_0\}$, where we use $J^{(1)}_0$ as the energy unit. In both cases $T^0_s<T'<T^0_d$, which means that the system is prepared in a TAP state. We will take the evolution Hamiltonian $H$ to be specified by $J=0.75 \, J^{(1)}_0$ in both cases. Now, using Eq.~(\ref{eq:final-energy}) it is easy to verify that both systems have the same energy density, $e_f=-0.325 \ J^{(1)}_0$. If the only conserved quantity is energy, that is the Hamiltonian $H$ itself, and both systems thermalise, the asymptotic states reached should be identical. However, the asymptotic states predicted by Eqs.~(\ref{eq:energy-simplified2})--(\ref{eq:q-simplified}), that we have checked against the full dynamics, are different $\{q^{(1)}=0.5, \ q_0^{(1)}=0.548, \ T_f^{(1)}=0.513 \, J^{(1)}_0\}$ and $\{q^{(1)}=0.529,\ q_0^{(1)}=0.578, \ T_f^{(1)}=0.519 \, J^{(1)}_0\}$. In particular, the final temperatures are different. We can interpret this fact using the results in Section~\ref{subsec:dyn_metastable}. Both systems are initialised in a TAP state and, since we inject only a small amount of energy, the subsequent dynamics takes place inside the same TAP state that is only translated in energy and rescaled in size by the quench. In other words, both systems are unable to forget the details of the initial condition, in particular which is the initial TAP state. Both systems reach thermal equilibrium inside the TAP state, however this bounded equilibrium is not compatible with thermal equilibrium in the whole phase space. In fact, for both quenches the final temperature $T^{(1,2)}_f$ is larger than the dynamical temperature $T_d=0.459 \, J_0^{(1)}$
and both systems show $q\neq0$. However, the equilibrium state described by the Gibbs measure $e^{-\beta_f H}/Z$ is paramagnetic for any
$T_f>T_d$, with $q=0$. This confirms that the asymptotic states reached by both systems are not thermal.

Finally, quenches ending up in a paramagnetic asymptotic state with $q=q_0=0$ show thermalisation. Let us consider two different initial conditions $(1)=\{J^{(1)}_0, \ T'^{(1)}=0.6 \, J^{(1)}_0\}$ and $(2)=\{J^{(2)}_0=1.05 \, J^{(1)}_0, \ T'^{(1)}=0.615 \, J^{(1)}_0\}$. Again, both initial conditions are prepared in a TAP state. Now we will consider a situation in which we inject a larger amount of energy by choosing $J=0.4 \, J^{(1)}_0$. In this case, both systems have the same energy density $e_f=-0.033 \ J^{(1)}_0$.
The asymptotic state in both cases is paramagnetic and the final temperatures are also equal $T_f^{(1)}=T_f^{(2)}=0.368 \, J^{(1)}_0$. Since the final temperature is larger than the dynamical temperature $T_d=0.244 \, J^{(1)}_0$, the asymptotic state is compatible with the paramagnetic equilibrium state $e^{-\beta_f H}/Z$.

\vspace{0.25cm}

We end with a discussion of the temperatures selected by the asymptotic state, be it a steady or an ageing one.
The steady state can be a simple paramagnet or a metastable TAP state. In both of these cases,
the model fixes its temperature, that is a function of the energy after the quench,
and the fluctuation dissipation theorem is satisfied with respect to it.
In the ageing asymptotic state, instead, the dynamics occur in two, well separated, time regimes controlled by the
relation between the time-delay and the reference or waiting time.
The fluctuation-dissipation relation is satisfied with respect to these
two temperatures for the time-delay taking values in the two regimes. The crossover between the two regimes
takes place when the correlation function passes by the plateau value $q$. The ageing regime has exactly the same features as
for the dissipative system.

It was recently shown that in quenches of isolated quantum {\it integrable} systems the fluctuation dissipation
relation, as a function of frequency, allows one to measure, in the steady state,
the Generalised Gibbs Ensemble effective temperatures,
one for each conserved quantity~\cite{FoGaKoCu16,deNardis-etal}. In the model we studied in this
paper there is only one conserved quantity, the total energy, but, in the ageing asymptotic state, the
system acquires two temperatures, depending on the range of frequencies at
which it is examined. Whether there is a link between the temperatures ($T_f$ and $T_{\rm eff}$)
measured in the out of equilibrium non-integrable systems and the ones of the GGE of integrable
models, is a question that deserves further analysis.

\vspace{0.25cm}

The TAP states are  separated by barriers diverging with the system size,
that is to say, with $N$ in this case.
In the dissipative problem the escape time from the TAP states is
expected to grow exponentially, {\`a la} Arrhenius, with the surrounding
(free-energy) barrier over the thermal energy $k_BT$ and, therefore, exponentially with $N$.
In the infinite size limit, the TAP states are fully confining even under thermal fluctuations.
Numerical evidence for the finite (though very long)
lifetime of the trapping states for finite $N$ was given in, e.g.,~\cite{CuKuLePe97,BeCuIg01,Be03}.

As shown by our results, in the isolated problem,  the asymptotic quench dynamics
of the infinite size system can also be confined to these states with special choices of the initial conditions (that are
correlated with the random potential). The calculation that we presented in this paper are valid only in the
infinite size limit. In the isolated case, though, we expect that the
remnants of the TAP states in the finite $N$ systems should be confining until some crossover energy input is
exceeded; how this is achieved should depend on the parameters.

Under these conditions, a system initially prepared in a configuration belonging to a finite $N$ pseudo-TAP
state will initially explore the region of the phase space corresponding to the transformed finite $N$ pseudo-TAP state,
before relaxing to the final paramagnetic state in a longer timescale. The crossover time should scale with $N$, becoming infinite for $N\rightarrow \infty$, in concordance with our results. This two step relaxation would be reminiscent of the situation in nearly integrable isolated quantum many body systems, which initially relax to a metastable state (called prethermalised state in the recent literature~\cite{KoWoEc11,NeIuCa14,BerEssGroRo15}) before reaching equilibrium in a longer time-scale~\cite{BerEssGroRo15}. Prethermalised states in nearly integrable systems are in correspondence with the non-thermal stationary states of the associated integrable model~\cite{KoWoEc11,NeIuCa14}. In this case, the crossover time-scale scales with the distance to integrability, and diverges for the integrable model, for which the non-thermal stationary state, often described by GGE density matrices, are the truly final stationary states.

\vspace{0.25cm}

In a separate publication we will discuss how the methods and results of this paper extend to the quantum problem.

\vspace{1cm}

\noindent
{\large {\bf Acknowledgements.}}
We warmly thank D. Abanin, G. Biroli, T. Grover, J. Kurchan, A. Tartaglia and M. Tarzia for very useful discussions.
LFC thanks the KITP Santa Barbara for hospitality during part of the preparation of this work.
We acknowledge   financial
support from ECOS-Sud A14E01, PICS 506691 (CNRS-CONICET Argentina) and
NSF under Grant No.  PHY11-25915. LFC is a member of
Institut Universitaire de France.

\bibliographystyle{phaip}

\end{document}